# Time Resolved Study of Laser-Induced Ultrafast Alloying Processes in Au/Pd Core–Shell Nanorods

Abhisakh Sarma*, Jayanath C. P. Koliyadu, Romain Letrun, Egor Sobolev, Trupthi Devaiah C, Agnieszka Wrona, Katerina Doerner, Diogo V. M. Melo, Marco Kloos, Huijong Han, Marcin Sikorski, Konstantin Kharitonov, Juncheng E, Joana Valerio, Pralay K. Santra, Erik M. J. Johansson, Richard Bean, Chan Kim*, Tokushi Sato*

**ABSTRACT:** Femtosecond laser-induced alloying presents a novel approach to modifying bimetallic systems. Visualizing ultrafast processes during laser-induced alloying is essential to uncover fundamental mechanisms associated with phase transformations, which enables precise control over material composition and structure at the atomic level. In this study, we investigated the ultrafast dynamics of laser-induced alloying of Au/Pd core-shell nanorods using a time-resolved X-ray diffraction technique at an X-ray free-electron laser facility, capturing the structural evolution from picoseconds to microsecond timescales. We found that a laser fluence threshold of ~ 48 mJ/cm$^2$ with 800 nm excitation is sufficient for melting and subsequent alloy formation. Above this threshold, the formation of $Au_{1.51}Pd_{0.49}$ was observed, and we found that alloying is not a single-step phenomenon; instead, it is a dynamic process involving interdiffusion.

## INTRODUCTION

The advent of nanomaterials has revolutionized various fields[1], from material science to medical applications, owing to their unique optical, electronic, and chemical properties[2,3]. Among these materials, core-shell nanostructures[4], such as gold (Au)/palladium (Pd) core-shell nanorods (NRs), have drawn attention mainly due to their enhanced stability, tunable properties[5], and potential applications in catalysis[6–8], drug delivery[9], and material detection[10–12]. The unique combination of the high electron density and the catalytic activity of these individual elements makes Au/Pd core-shell NRs particularly promising for technological advancements[3,13–15].

The alloying of Au/Pd nanostructures can lead to tunable properties[5] that are highly desirable for various applications. However, the precise control of the alloying process within these core-shell NRs remains a significant challenge[16–20]. The traditional methods—such as thermal annealing, chemical reduction, and chemical vapor deposition—often involve prolonged processing times, high temperatures, or slow diffusion kinetics, which may not guarantee uniformity or precise control of the structure at an atomic level.

One of the most traditional approaches to alloying in core-shell structures is thermal annealing to drive atomic diffusion between the core and the shell[21–24]. This method often suffers from limitations in controlling the extent and uniformity of alloying, particularly for nanostructures. High temperatures can also lead to significant structural damage or phase separation[23]. A key challenge of this technique is maintaining the desired shape during the alloying process in a dry environment, because the increased atomic diffusion at high temperature can cause the metals to melt and merge. Without an extra protective layer, such as silica, the nanoparticles can easily undergo sintering and reshaping during the alloying process, resulting in the loss of their intended functionality[23]. Another method to fabricate alloy NRs starting from core-shell nanostructure is the Chemical Vapor Deposition (CVD) technique. The CVD technique allows the deposition of metal layers on core-shell nanostructures, which can undergo alloying by adjusting the deposition conditions. This method provides a degree of control over the thickness and composition of the shell, although it is still limited by diffusion kinetics.

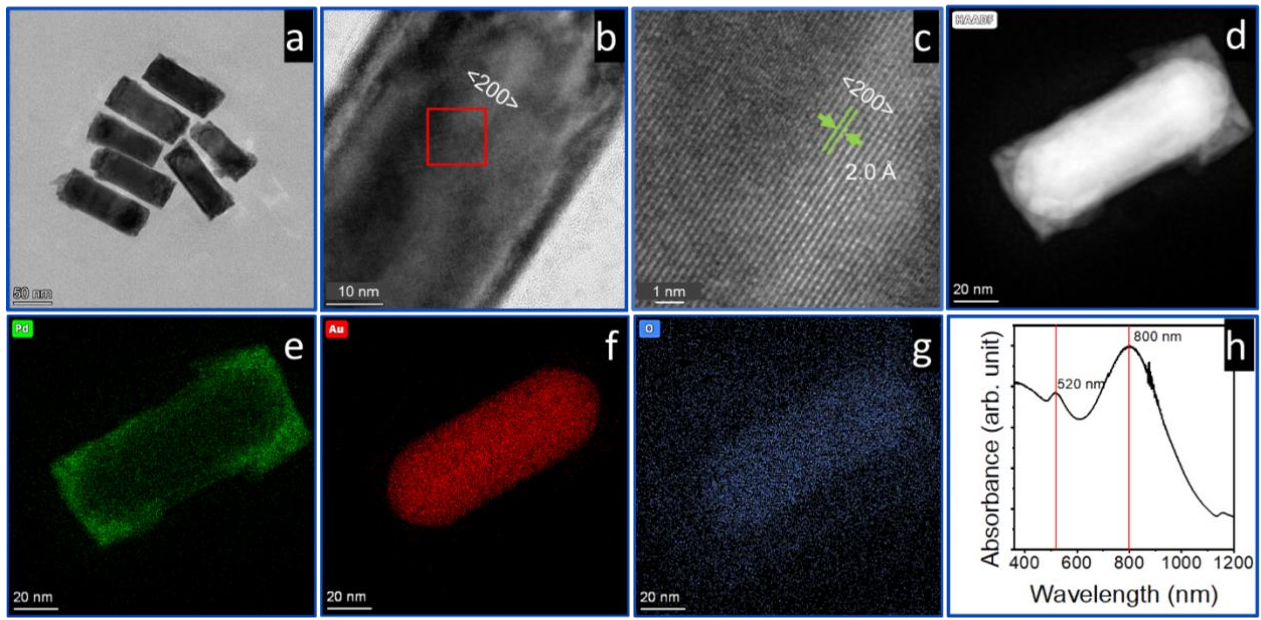

Figure 1: (a) TEM image of synthesized nanorods. (b) HRTEM image showing detailed structural features. (c) Lattice-resolved HRTEM image of the region marked by red box in (b), indicating an interplanar spacing of 2.0 Å, corresponding to the Au (200) plane. (d) High-angle annular dark-field scanning TEM (HAADF-STEM) image highlighting the contrast variations. (e–g) EDS elemental maps showing the distribution of (e) palladium (green), (f) gold (red), and (g) an additional oxygen element (blue). (h) UV–vis–NIR absorption spectrum displaying peaks at 520 nm and 800 nm, characteristic of plasmonic resonance in minor and major axes, respectively.

Recent advancements in femtosecond laser technology present promising opportunities to overcome these limitations by inducing alloying at nanometer length scales[25,26]. This approach offers a unique capability since it can provide extremely fast localized heating with minimal thermal diffusion into the surrounding materials. Therefore, femtosecond laser pulses can induce rapid melting and mixing of the core and shell materials without significantly affecting the overall structure[25,26], making it an ideal technique for alloying. Laser irradiation also allows for the tuning of alloying depth and selective modification of the nanostructure by varying laser wavelength and pulse duration. Despite the growing interest in femtosecond laser-induced alloying methods, there is a notable gap in the literature regarding *in situ* studies that capture the real-time dynamics of alloying processes in core-shell NRs. Current research has largely focused on theoretical predictions or post-treatment characterizations, leaving a critical need for experimental insights into the fast, transient processes that govern alloying at the nanoscale[25–29].

In this study, we investigate the femtosecond laser-induced alloying of Au/Pd core–shell NRs by capturing their structural evolution at ultrafast timescales (ps–μs) using time-resolved X-ray diffraction (TR-XRD). Various laser fluences were applied to explore how these energetic inputs influence the structural and compositional transformations. A key innovation of our work is the use of a single optical pulse excitation and probing strategy. Unlike traditional multi-pulse excitation methods[25–28,30–32], which can introduce cumulative effects or alter material properties due to repeated laser pulses, our approach ensures that each nanoparticle is exposed to only one laser pulse. This distinctive single-shot design eliminates the confounding effects associated with cumulative structural changes or irreversible damages that can occur under repeated laser exposure. By ensuring that each NR is only exposed to a single ultrafast excitation event, we capture the intrinsic, non-averaged structural dynamics of the Au/Pd core–shell NRs with high temporal fidelity. Our approach yields insights into the fundamental dynamics of alloying and phase transitions at the nanoscale where the effects of multi-pulse exposure may be obscured.

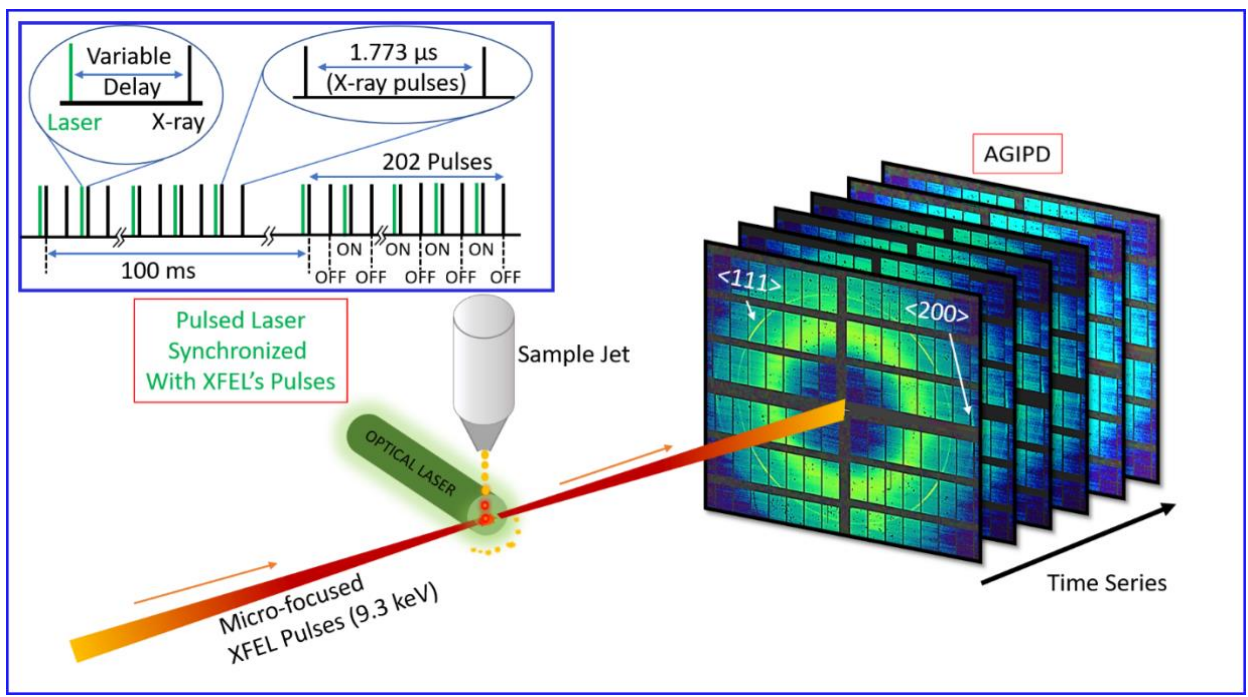

Figure 2: A schematic of the experimental setup for TR-XRD. Optical laser pulses are synchronized with the XFEL pulses to irradiate a sample jet. The time-series diffraction patterns were collected using the Adaptive Gain Integrating Pixel Detector (AGIPD). (Top left inset) 10 Hz burst repetition of EuXFEL's pulses, where each 100 ms burst contains 202 X-ray pulses and the pulse-pulse separation is 1.773 μs (564 kHz). The timing scheme of the laser and X-ray pulses, showing optical laser pulses (green colored pulses) on the odd-number of X-ray pulses (black colored pulses) in the train. The time delay between laser and X-ray pulses were varying from ps-μs.

## RESULTS AND DISCUSSIONS

### Morphological and Optical Properties of Pristine Nanorods.

To establish a baseline for understanding the laser-induced dynamics, we first characterized the pristine Au/Pd core–shell NRs using transmission electron microscopy (TEM), energy-dispersive X-ray spectroscopy (EDS), and optical absorption spectroscopy. These analyses provided key structural, compositional, and optical insights which are essential for interpreting their transient behavior under femtosecond laser irradiation. Figure 1a shows a TEM image of the core-shell NRs with an average aspect ratio of ~2.4, a diameter of 44 ± 4 nm, and a length of 106 ± 12 nm (size distribution in Figure S1). High-resolution TEM (HRTEM) images revealed that the <200> lattice planes in the gold core are aligned along the minor axis of the NRs. Notably, these planes in the face-centered cubic (fcc) structure typically lie along the rod length as well. Additional HRTEM images are shown in Figure S2 (section SI-1). EDS elemental maps (Figures 1e–1g) confirm the core–shell architecture, with gold forming a cylindrical core and palladium preferentially accumulating at the NR ends. Oxygen was predominantly localized in the Pd-rich shell, suggesting partial oxidation (Figure 1g). UV–vis–NIR absorption spectrum (Figure 1h) displays two distinct peaks: one at ~520 nm, which corresponds to lateral plasmon resonance, and another at ~800 nm represents the longitudinal resonance absorption. These optical signatures validate the NRs' morphology and play a pivotal role in their laser-driven photothermal dynamics.

**Ultrafast Laser-Induced Alloying Dynamics.**

The TR-XRD experiment was carried out at the Single Particles, Clusters, and Biomolecules and Serial Femtosecond Crystallography (SPB/SFX) instrument[33] at the European X-ray Free-Electron Laser (EuXFEL) facility. A schematic of the experimental setup is shown in Figure 2. The Au/Pd core–shell NRs were delivered by a water jet using a Gas Dynamic Virtual Nozzle (GDVN)[34]. During injection, the NRs exhibited a preferential alignment parallel to the jet flow direction, driven by their anisotropic shape and hydrodynamic interactions[35,36]. This alignment, which influences the polarization-dependent optical absorption, is critical for accurately interpreting the photothermal response (see section SI-2 for alignment calculations). The Experimental Details section provides further details on the experimental parameters, sample preparation, and optical pump laser.

Using TR-XRD, we tracked the structural changes of the Au/Pd core-shell NRs as a function of laser fluence across time delays from picoseconds to microseconds. The evolution of lattice constants derived from Bragg peaks of Au and Au/Pd alloys was analyzed and is shown in Figure 3. When the system is excited with a femtosecond laser pulse, it undergoes these paths: (i) ultrafast expansion due to electron–phonon (e–ph) coupling (≤10 ps), (ii) thermal expansion and heat dissipation by phonon–phonon (ph–ph) coupling (50–500 ps), and (iii) either recovery to the original state (low laser fluence) or alloy formation (high laser fluence) (500 ps to 1 µs).

**Electron–Phonon Coupling and Lattice Expansion (≤10 ps):**

Immediately following laser excitation (≤10 ps), a rapid expansion of the Au⟨111⟩ lattice driven by e–ph coupling[37–39] is observed (Figure 3a). The absorbed optical laser energy is initially deposited into the electronic subsystem, which rapidly transfers to the lattice, resulting in ultrafast thermal expansion. The magnitude of the lattice expansion increases as the laser fluence increases, which is consistent with the increased photon absorption. Importantly, this expansion remains linear at low fluences (≤20 mJ/cm²) and is reversible, as shown in Figure 3. These early dynamics represent the initial heating phase of the NRs.

**Phonon–Phonon Coupling (50–500 ps):**

At time delays between 50 ps and 500 ps, the system enters a relaxation phase and reaches thermal equilibrium, where ph–ph coupling becomes dominant. In this period, thermal energy begins to dissipate from the gold lattice into the surrounding water medium through the palladium shell. This heat dissipation is evidenced by a noticeable decrease in lattice expansion after ~ 100 ps at low laser fluences (≤20 mJ/cm²) and after ~ 500 ps at high laser fluences (100 mJ/cm² and 200 mJ/cm²), as shown in Figure 3a. The lattice expansion reaches a peak at around 100 ps for low laser fluences and then cools down due to environmental heat exchange, causing contraction of the lattice. This behavior deviates at higher laser fluences (100 mJ/cm² and 200 mJ/cm²), where the lattice expansion continues to increase until 500 ps and then relaxes afterward, consistent with a delay in energy dissipation due to higher initial heating.

Beyond the expansion-related effects, an asymmetric broadening of the Au⟨111⟩ Bragg peak was observed (Figure S7). This feature indicates inhomogeneous lattice expansion, which is presumably due to transient thermal gradients within the NRs. These gradients can originate either during the heating or develop during the cooling of the nanostructures. As previously mentioned, the lattice expansion reaches its maximum value within a time delay range of 100 ps (at lower fluences) to 500 ps (at higher fluences). Moreover, the good agreement between the reduction in the Au⟨111⟩ Bragg peak intensity and the Debye–Waller factor (discussed in detail in the following section) indicates that the system reaches thermal equilibrium within this timeframe during the heating process. In addition, careful observation of the Au⟨111⟩ Bragg reflection (Figure S7), even at low fluence, clearly reveals the onset of pronounced asymmetry starting at approximately 500 ps. This suggests that asymmetric peak broadening becomes dominant at the beginning of the cooling phase. Such behavior reflects spatially nonuniform cooling across the anisotropic Au/Pd core–shell NRs, leading to nonuniform lattice strain.

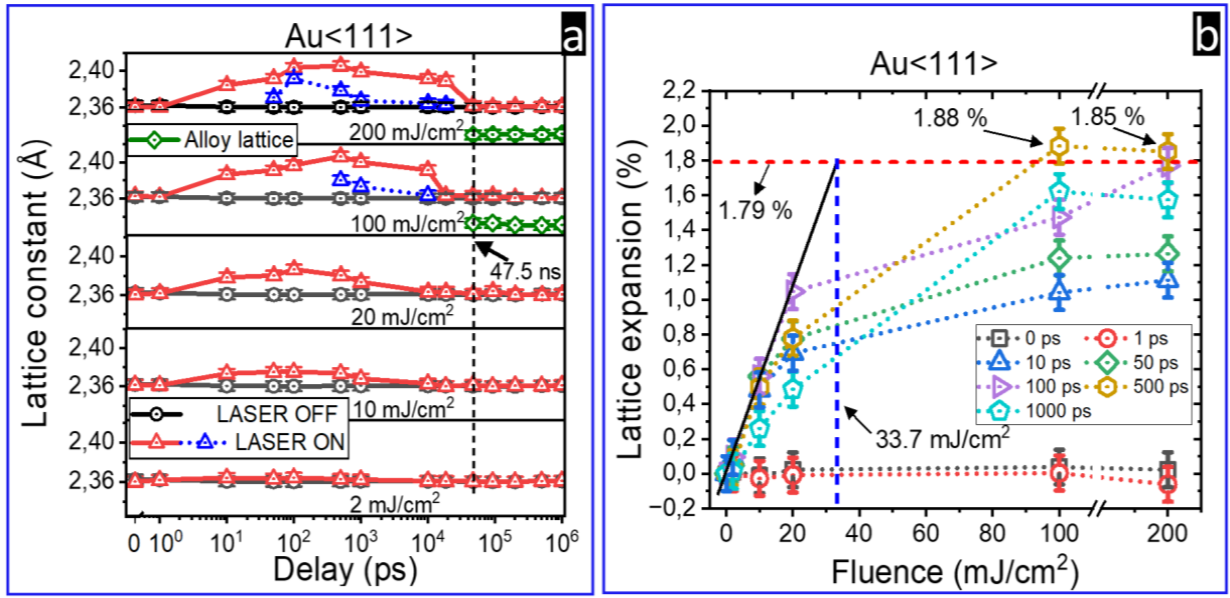

Figure 3: Ultrafast lattice expansion and relaxation dynamics of Au<111> under varying laser fluences. (a) Temporal evolution of the Au<111> lattice constant for different laser fluences (2–200 mJ/cm²). The red, blue, and green data points represent the laser-on condition, while the black data points correspond to the laser-off condition. The green diamond markers indicate the formation of a new alloy phase ($Au_{1.51}Pd_{0.49}$) observed at 100 mJ/cm² and 200 mJ/cm². The two distinct sets of red and blue data points at these fluences reflect the splitting of Bragg <111> reflections within the delay time range of 50 ps to 20 ns. The vertical dashed line at 47.5 ns marks the delay time after which the alloy reflection is visible. A scale break at 0.5 ps on the x-axis was applied. The section before the break is linear, while the section after the break follows a logarithmic scale. (b) Lattice expansion (%) as a function of laser fluence in different time delays (0 ps to 1000 ps). The horizontal dashed red line represents the maximum expansion limit of gold before melting, the black solid line with a slope represents the linear fitted extrapolated line of the 100 ps data, and the vertical dashed blue line indicates the intersection point between the black solid and red dashed lines. A scale break in the Fluence axis from 110 mJ/cm² – 180 mJ/cm² was applied.

As these thermal gradients are relaxed through heat diffusion into the surrounding water, the diffraction peaks gradually become symmetrical. At later time delays (>500 ps), the asymmetric peak shape has started to recover to a symmetric profile, indicating that the lattice has initiated a return to a more thermally homogeneous state. The absence of these features in laser-off reference data further confirms that the transient peak broadening and the asymmetric peaks are induced by ultrafast laser excitation and not due to static structural inhomogeneities or measurement artifacts. To further ensure that the observed changes were thermally driven and not due to non-equilibrium or field-enhanced reshaping effects, two specific design considerations were implemented. The stretched pulse duration (890 fs) was deliberately chosen to reduce the peak electric field and thereby suppress non-thermal deformation mechanisms. This design was guided by the work of Plech *et al.*[40], who demonstrated that sub-200 fs pulses can drive anisotropic reshaping via near-field effects and electronic instabilities, while longer pulses favor energy transfer to the lattice and thermal expansion. Our use of stretched pulses thus ensured that the observed lattice alloying arises primarily from thermally driven dynamics, rather than impulsive or non-equilibrium effects.

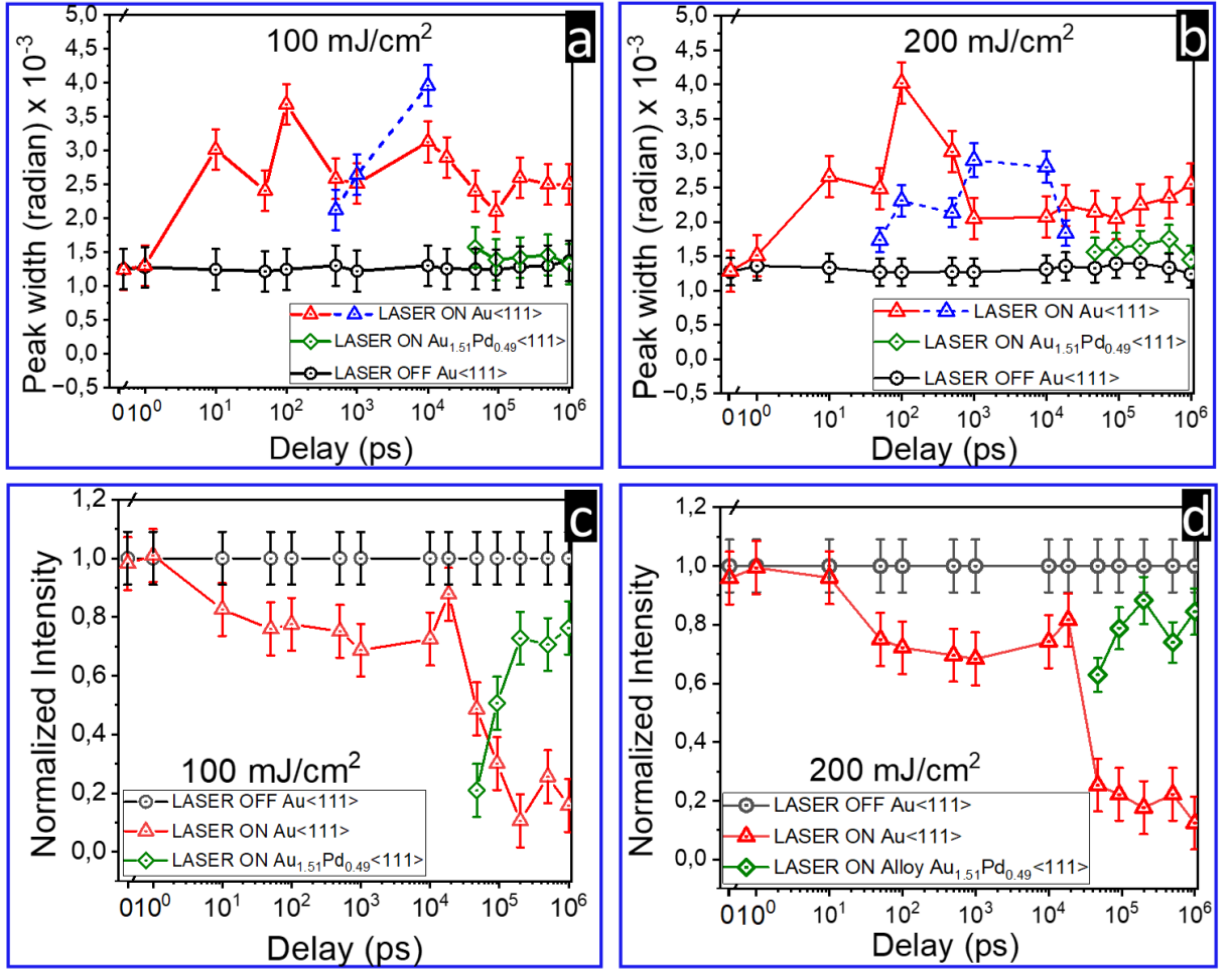

Figure 4: (a–b) Temporal evolution of the full width at half maximum (FWHM) of the Au⟨111⟩ peak for laser fluences of 100 mJ/cm² and 200 mJ/cm², respectively. Peak broadening indicates nonuniform lattice strain and transient disorder. Blue triangles highlight the second diffraction peak arising from peak splitting during transient alloying. (c–d) Normalized intensities of Au⟨111⟩ reflection for 100 mJ/cm² and 200 mJ/cm² fluences. Red points denote laser-on data, black points represent laser-off baselines, and green markers correspond to the $Au_{1.51}Pd_{0.49}$ alloy contribution.

In addition, we used circularly polarized excitation to eliminate orientation-dependent absorption effects. In systems with anisotropic NRs—particularly under conditions of partial alignment (see SI-2), as in our liquid jet—linearly polarized light can induce directionally biased reshaping, as previously reported by González-Rubio *et al.*[41]. Circular polarization averages out these anisotropies and promotes more uniform energy deposition across the ensemble. Together, these design choices provide a controlled excitation regime in which single-pulse thermal thresholds can be reliably probed.

**Relaxation Dynamics of the System:**

As described above, the lattice undergoes reversible expansion and starts relaxation around 100 ps at low laser fluences (≤20 mJ/cm²), indicative of ph–ph mediated heat dissipation. In parallel with the structural contraction, a decrease in diffraction intensity was observed (see Figure 4, S7 and S8), which provides a measure of lattice disorder caused by incoherent thermal motion or melting of the atoms. The integrated peak intensity ratios of the measured Au⟨111⟩ Bragg reflection between the laser on and off data at laser fluences of 10 mJ/cm² and 20 mJ/cm² were 75% ± 9% and 71% ± 9%, respectively. To determine the contribution of the Debye-Waller factor to the reduction in scattering intensity, the temperature of the nanoparticles must be estimated first. The lattice expansion data measured at 100 ps time delay was used to estimate the temperature since the system reaches a quasi-equilibrium state at this time delay. For comparison, we plotted the evolution of the lattice expansion as a function of the laser fluence for different time delays and observed a region of maximum linearity at 100 ps time delay in the low laser fluence (≤20 mJ/cm²), as indicated by the black solid line in Figure 3b. A similar linear behavior of lattice expansion was previously reported by Plech et al.[42] and the authors claimed that this slope is caused by the temperature-dependent increase in the coefficient of volume expansion and specific heat at lower laser fluence cases.

Assuming linear thermal expansion behavior, the lattice expansion ($\Delta a/a$) is related to the temperature rise ($\Delta T$) via:

$$\frac{da}{a} = \alpha(T)dT$$

where $\alpha(T)$ is the temperature-dependent thermal expansion coefficient. By using literature values for the thermal expansion coefficient of gold[43] and the measured lattice strain at 100 ps delay, we estimated the corresponding lattice temperatures as a function of the laser fluence (see section SI-3 for detailed calculations). The yielding lattice temperatures at low laser fluences (10 mJ/cm² and 20 mJ/cm²) are approximately 670 K and 950 K, respectively. The rather linear temperature-dependent volume expansion suggests that photon absorption is efficient and predictable in this regime, and the system remains below the melting threshold. Ultimately, to estimate the optical absorption cross-section of the Au/Pd NRs under experimental conditions, we used the measured lattice expansion to calculate the equilibrium lattice temperature and back-calculated the absorbed energy per particle using known thermophysical properties. This yielded a value of approximately 2.13 × 10⁻¹⁵ m² (Section SI-3, Figure S6), corresponding to the specific fluence, beam spot size, and single-pulse geometry of the XFEL experiment. Although the UV–vis extinction spectrum (Figure 1h) captures the overall spectral profile of the NRs, no absolute absorption cross-section was derived from this measurement. While absolute calibration is technically feasible using well-characterized cuvettes and particle concentrations, such a calibration would not reflect the conditions of the XFEL experiment, including single-pulse excitation, circular polarization, and partial alignment in the liquid jet. Therefore, we relied on temperature-based modeling of the absorbed energy to extract an effective absorption cross-section relevant to our actual experimental geometry

Notably, González-Rubio et al.[41] highlighted similar challenges, stating that random NR orientation and aspect ratio polydispersity in solution prevented reliable estimation of absorbed energy per particle from optical measurements. In contrast, our experiment minimizes these limitations by using a flowing liquid jet with partial NR alignment (confirmed in SI-2), circular polarization to average orientation effects, and TR-XRD to directly probe structural responses. Although absorption cross-section data for Au/Pd core–shell NRs are scarce, the value we obtained aligns well with literature values for pure Au NRs of similar aspect ratio (~1.87 × 10⁻¹⁵ m², Cho et al.[44]). This consistency supports the validity of our temperature-based approach, even in the absence of absolute optical calibration. This linear dependence between lattice expansion and laser fluence also provides a foundation to estimate the laser fluence required to induce structural phase transitions.

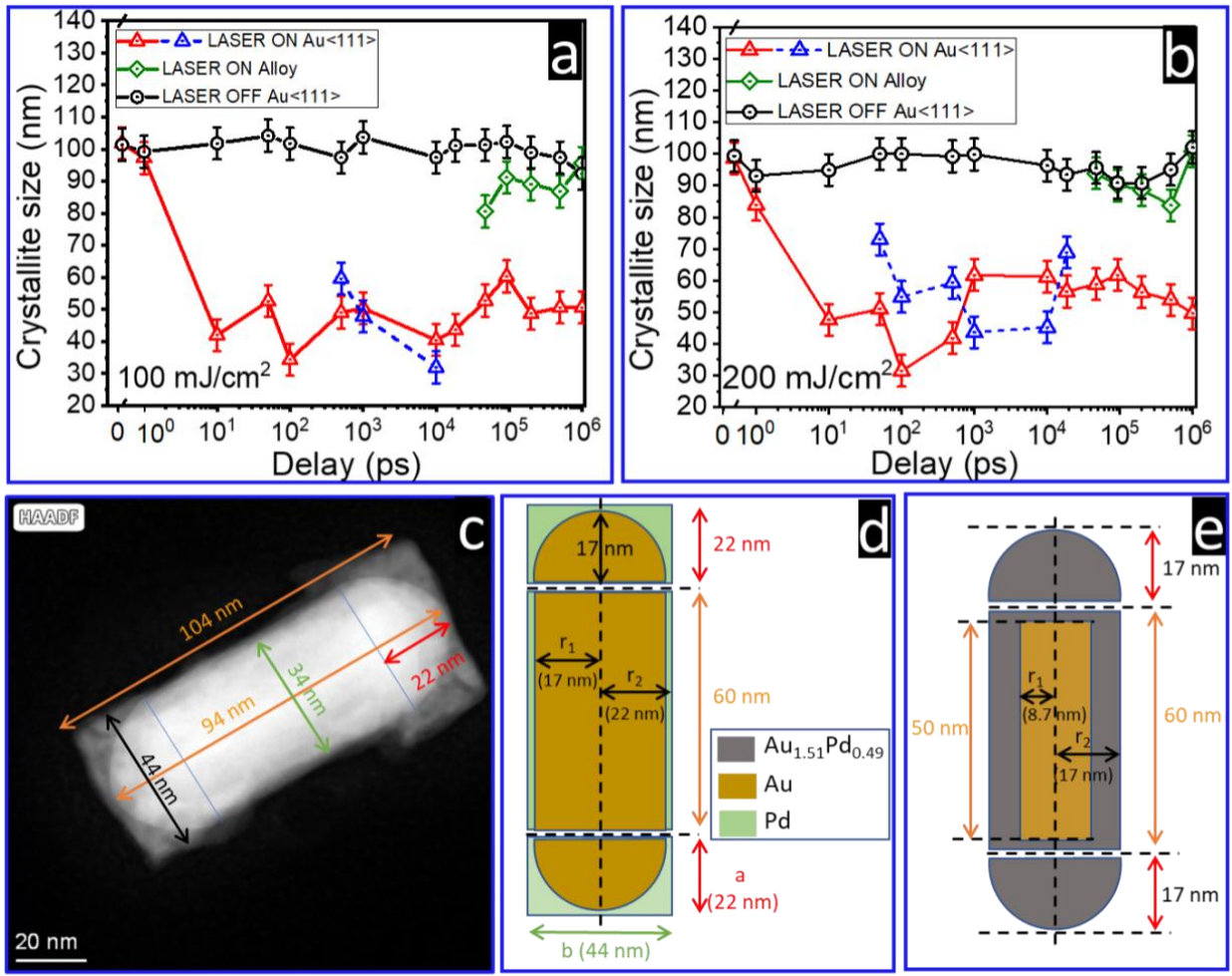

Figure 5: Structural characterization of the alloyed Au/Pd NRs. Crystallite size of Au⟨111⟩ and $Au_{1.51}Pd_{0.49}$ phase as a function of time-delay for laser fluences of (a) 100 mJ/cm² and (b) 200 mJ/cm², calculated using the Scherrer equation from TR-XRD peak widths. (c) HAADF-STEM image of a pristine Au/Pd core–shell NRs with a measured length of ~ 104 nm. (d) Schematic of the initial structure prior to laser excitation, showing a central gold core surrounded by a palladium shell. (e) Schematic model of the final alloyed structure after femtosecond laser irradiation, showing partial retention of the gold core and formation of an $Au_{1.51}Pd_{0.49}$ alloy phase at the periphery. Dimensional estimates are based on TR-XRD derived volume fractions and crystallite sizes (200 mJ/cm² at 1 μs time delay).

The predicted decreases in Bragg intensity at 670 K and 950 K, corresponding to the laser fluences of 10 mJ/cm² and 20 mJ/cm², are 87% and 79%, respectively, and these predictions agree well with the experimental results, which are 75% ± 9% and 71% ± 9%, as mentioned earlier, although the error bars are rather large. The agreement between experiment and theory confirms that, at these laser fluences, the system remains below the melting threshold and the observed intensity loss is purely a result of thermal disorder rather than structural phase transitions. Although no significant melting is expected in this regime, it is important to note that surface melting effects may begin to contribute at intermediate fluences. Prior studies on metallic nanoparticles have shown that surface atoms can exhibit reduced thermal stability and begin to disorder at temperatures below the bulk melting point[45].

The observed relaxation in the lattice expansion (from contraction of the unit cell) and the Debye–Waller-based reduction in diffraction peak intensity provide complementary evidence of phonon-mediated thermal relaxation. The decrease in lattice constant after lattice expansion indicates that the NRs are cooling down and returning to their equilibrium as thermal energy dissipates into the surrounding medium. At the same time, the reduced diffraction intensity reflects increased thermal motion of the atoms, which is a hallmark of vibrational disorder, but not structural breakdown.

To determine the laser fluence required to initiate melting in the Au/Pd core–shell NRs, we combined experimental lattice expansion data (lattice temperature) with theoretical enthalpy considerations. We extrapolated the linear increase in lattice expansion as a function of laser fluence (black solid line in Figure 3b) to intersect the 1.79% expansion mark (horizontal dashed red line), known to be the bulk melting threshold for gold, and this intersection corresponds to a laser fluence of 33.7 mJ/cm².[42] This suggests that the laser fluence of 33.7 mJ/cm² is sufficient to raise the gold core to its melting point with 800 nm excitation. However, the latent heat of fusion must be overcome for complete melting. Based on literature[46,47], the enthalpy required to raise bulk gold from room temperature to its melting point is approximately 29.1 kJ/mol, and an additional 12.7 kJ/mol of energy is required to complete the melting. The latent heat thus accounts for ~ 43.7% additional energy input. Assuming a linear relationship between laser fluence and absorbed energy, we estimate that an extra 14.7 mJ/cm² is required to fully melt the gold NRs, yielding a total theoretical melting fluence of approximately 48.4 mJ/cm².

To verify the complete melting of the Au/Pd core-shell NRs, we performed measurements at laser fluences of 100 mJ/cm² and 200 mJ/cm², above the estimated melting threshold. At these fluences, the lattice expansion exceeds 1.79% at 500 ps (Figure 3b), and it shows strong peak broadening and intensity loss (Figures 4a and 4b) in the Au⟨111⟩ Bragg peak. At 500 ps time delay, the integrated peak intensity ratios of the measured Au⟨111⟩ Bragg reflection between the laser on and off data at the laser fluences of 100 mJ/cm² and 200 mJ/cm² are 75% ± 9% and 69% ± 9%, respectively. The calculated contribution of the Debye-Waller factor at the melting temperature of gold (1330 K) is around 68% (section SI-6). Even at high laser fluences (100 mJ/cm² and 200 mJ/cm²), no significant signature of phase transition or melting of Au was observed at this time delay. Strong asymmetric peak broadening and expansion of the lattice constant (1.88 % and 1.85 %) were observed, beyond the maximum expansion limit of gold before melting (1.79 %), corresponding to a superheated state. A similar behavior of superheated spherical gold nanoparticles was previously reported by Plech *et al*[42].

The persistence of well-defined Bragg peaks, even as intensity drops and crystal size decreases (Figures 5a and 5b), confirms that long-range crystalline order is maintained throughout this time window (50 ps – 500 ps). It indicates that the NRs are in a dynamically disordered crystalline state: atoms vibrate more vigorously due to the absorbed energy, without the atomic diffusion or lattice collapse associated with melting. This interpretation is consistent with the low laser fluence regime, where the excitation energy is sufficient to heat the lattice but insufficient to drive irreversible phase transitions or alloy formation.

**Onset and Stabilization of Alloy Formation (10 ns–1 μs):**

Beyond the superheated crystalline state observed at the high laser fluences (100 mJ/cm²and 200 mJ/cm²), we track the formation of a new alloy phase over nanoseconds to microseconds after laser irradiation. A new Bragg reflection at q ≈ 2.696 Å⁻¹ was observed around 47 ns time delay, corresponding to the $Au_{1.51}Pd_{0.49}$ alloy phase (Figure 3a and S7). The appearance of the alloying peak indicates the onset of interdiffusion between the gold core and the palladium shell, which is triggered by the transient melting or highly disordered states formed at the earlier time delay. Between 47 ns and 200 ns, the intensity of this alloy peak continues to increase, indicating domain growth of the alloy. Along with this, the original Au⟨111⟩ peak persists with a decrease in intensity, indicating a partial transformation of the gold into the new phase. These observations confirm that alloying begins on nanosecond timescales following energy deposition and transient lattice disorder.

To gain a comprehensive understanding of the final structure and enable quantitative analysis, the diffraction signal at longer time delays (100 ns to 1 μs) was analyzed. At these time delays, the position and width of the alloy peaks stabilize, confirming that the structure has reached a thermodynamic steady state. This behavior reflects that the atomic interdiffusion and recrystallization have been completed. By using the Scherrer equation[48], crystal size was extracted from the diffraction peak widths (Figure 5), and it was confirmed that the alloyed domains grow to approximately 100 nm along the ⟨111⟩ direction, which is comparable with the full length of the original NRs (Figure 5c). In contrast, the residual Au⟨111⟩ domains shrink to ~50 nm (~53 nm for 100 mJ/cm$^2$ and ~50 nm for 200 mJ/cm$^2$), suggesting that alloying predominantly occurs at the shell and outer core regions. These values indicate partial alloying, consistent with a mechanism involving interdiffusion rather than complete core dissolution.

To place these findings in context, we compared (see SI-8) our results with previous femtosecond laser studies on gold-based plasmonic nanostructures, which often report reshaping or melting at significantly lower fluences. Based on the calibrated relationship between lattice expansion and temperature (Figure S5b), we estimate that the lattice melting point corresponds to a fluence of approximately 48 mJ/cm². Our TR-XRD measurements, performed at higher fluences of 100 and 200 mJ/cm², reveal Bragg peak broadening and shifting consistent with Au–Pd lattice alloying. Importantly, even at these elevated fluences, no major morphological reshaping was observed, as confirmed by preserved diffraction symmetry and representative raw detector images.

Previous studies have reported tip rounding, reshaping, or aspect ratio changes of Au or Au@Pd nanorods at significantly lower fluences, particularly under multipulse conditions (e.g., González-Rubio et al.[41]; Zijlstra et al.[49]; Nazemi et al.[50]; Manzaneda-González et al.[51]). These transformations typically occurred in pure Au or alloyed shells under stationary conditions, where cumulative heating and strong surface atom mobility favored reshaping. In contrast, under our single-pulse excitation and X-ray probing conditions, Au/Pd core–shell nanorods exhibit greater resistance to large-scale aspect ratio changes. A possible explanation is that the Au core (melting point 1064 °C) may undergo transient melting, while the Pd shell (melting point 1555 °C, higher cohesive energy, slower diffusion kinetics) plausibly suppresses rapid surface deformation. Additional stabilization may arise from lattice mismatch at the Au–Pd interface, which increases the barrier for surface diffusion. Also, the dominant irreversible process detectable by TR-XRD is lattice-level Au–Pd interdiffusion, rather than major reshaping.

Although the alloying threshold fluence observed here (~48 mJ/cm²) is about an order of magnitude higher than reported in earlier Au nanostructure studies[41,42,49], normalization by absorbed energy per atom reduces this discrepancy substantially. At threshold, our Au/Pd nanorods absorb maximum ~0.57 eV per atom, compared to ~0.37 eV per atom for 44 nm Au spheres studied by Plech et al.[42] —a difference of only ~1.5 times. This suggests that fluence alone is not a reliable benchmark, as differences in particle geometry, atom count, and material stability strongly influence absolute thresholds. Instead, absorbed energy per atom provides a more robust metric, and in our case, plausibly explains why lattice-level alloying emerges as the prevailing transformation pathway under single-pulse excitation. In addition to the use of circular polarized lasers, it can minimize the nanorods' orientational absorption bias (see Optical Pump and XFEL Probe Geometry sub-section, in experimental details section), which might influence the threshold difference compared to the threshold result obtained using a plane polarized laser, as mentioned above.

Figures 5d and 5e schematically illustrate the NR structure before and after laser-induced alloying at the laser fluence of 200 mJ/cm². These schematics are constructed based on quantitative estimates derived from TR-XRD data, including crystallite size differences (Figure 5b), residual gold fractions (Figures 4d), and the known geometry of the pristine NRs (Figure 5c). By integrating these structural parameters, we developed a representative model illustrating how interdiffusion proceeds from the outer palladium shell to the interior, forming an alloy phase while retaining some of the original gold core. The model incorporates stoichiometric considerations for the data collected for both laser fluences (100 mJ/cm² and 200 mJ/cm²). For instance, volumetric analysis at the laser fluence of 100 mJ/cm² based on the residual Au⟨111⟩ intensity ratios suggest that the final gold core has a length and diameter of ~ 53 nm and ~ 19.6 nm, with a residual volume of ~ 16%, in good agreement with the measured intensity loss (Figure 4c). At the laser fluence of 200 mJ/cm², the core length reduces to ~ 50 nm and the corresponding diameter is reduced to ~ 17.4 nm, yielding a residual gold volume of ~ 12% (Figure 4d). This volume-based analysis (Section SI-5, Figure S9) also shows that the amount of gold consumed to form the $Au_{1.51}Pd_{0.49}$ alloy is ~ 7.1 times larger than the volume of palladium, consistent with the stoichiometry of the final product. These ratios are essential to understanding the mass redistribution and structural transformation during the alloying process. The partial retention of gold supports a model in which the pristine core–shell morphology is structurally preserved, but the internal composition is transformed.

The use of a rapidly flowing 4.9 μm water jet in combination with a high repetition-rate femtosecond laser ensured that each Au/Pd NR experienced only a single excitation event under well-defined conditions. This setup was essential for isolating ultrafast lattice responses but made it technically infeasible to collect irradiated particles for post-mortem imaging. Moreover, any *ex situ* analysis performed in static systems (e.g., cuvettes or drop-cast films) would fail to replicate key experimental parameters—namely, flow-induced alignment, thermal dissipation, and strict single-pulse exposure—and would thus not reflect the intrinsic alloying behavior we report. We also note that exciting a static droplet with a single optical pulse would not reproduce the conditions of our experiment. The optical spot (~40 μm) would irradiate nanorods with random orientations, leading to strongly heterogeneous excitation and cooling dynamics. As discussed in the Methods and SI-2.3, the flowing jet geometry provides partial flow-induced alignment (~69%) combined with circular polarization, minimizing such anisotropy. In contrast, TEM/EDX analysis of a dried droplet would risk artifacts dominated by unexcited rods and preparation effects. While microfluidic platforms may eventually allow single-pulse exposure with sample recovery, such approaches are beyond the scope of this study. Instead, we rely on *in situ* TR-XRD as a direct probe of lattice-level changes. The fluence-dependent lattice expansion, irreversible Bragg peak broadening, and diffraction asymmetry together indicate a permanent structural transformation consistent with alloying. While not atomically resolved, these features are consistent with phase change and interdiffusion as established in prior ultrafast diffraction studies.

The results show that single-pulse femtosecond laser irradiation can induce an interdiffusion process to produce Au/Pd alloy NRs, with nearly pristine morphology, but modified internal composition and domain structure. This approach demonstrates a pathway toward post-synthetic, light-driven tuning of multi-metallic nanomaterials with nanometer-scale precision. This mechanism highlights the potential of femtosecond laser pulses for post-synthetic phase engineering in bimetallic nanomaterials with excellent spatial and temporal control.

## CONCLUSION

This study establishes femtosecond laser-induced alloying as a versatile and precise method for tailoring the structural and compositional properties of bimetallic nanomaterials. TR-XRD at an XFEL facility in conjunction with a single-shot optical excitation strategy was utilized to directly capture the lattice evolution of Au/Pd core–shell NRs on picosecond to microsecond time scales. This single-pulse approach eliminates cumulative heating artifacts and enables us to observe transient structural changes during alloy formation in real-time.

Our results show that alloying is not a one-step phase transformation but a dynamic process governed by interdiffusion between the gold core and the palladium shell. We identified a fluence threshold of ~ 48 mJ/cm² required to fully melt the NRs and form a stable $Au_{1.51}Pd_{0.49}$ alloy, while the lattice expansion remains reversible below this threshold. Notably, even above the melting threshold, diffraction signals from superheated gold crystallites persist, suggesting the presence of metastable intermediate states during phase evolution. Quantitative analysis of the crystallite sizes before and after alloying reveals a reduction in gold domain size, supporting a partial preservation of the core morphology. This behavior highlights interdiffusion, rather than total dissolution, as the dominant alloying mechanism. The volume mismatch between the core and shell materials further emphasizes the complexity of nanoscale mixing under ultrafast heating and cooling processes. We believe that this work advances the fundamental understanding of femtosecond laser-driven alloying and provides a powerful framework for designing plasmonic and catalytic nanomaterials with tunable functionality.

Finally, while the *in situ* nature of our experiment provided high temporal resolution and tight control over single-pulse excitation, it also precluded direct post-irradiation imaging of the alloyed NRs. Attempts to replicate the experiment in a static or batch setup would compromise essential conditions such as NR alignment, thermal dissipation, and single-pulse exposure. Thus, the present interpretation of alloying is based solely on real-time diffraction evidence. Future work will focus on developing experimental platforms that maintain XFEL-equivalent conditions while enabling sample recovery for *ex situ* morphological validation. Our single-pulse approach using ultrashort XFEL pulses offers new opportunities for investigating light–matter interactions in nanoscale systems with atomic-scale precision.

## EXPERIMENTAL DETAILS

**Sample & Characterization.** The Au/Pd core-shell NRs (part number Pd-NR-250) purchased from NanoSeedz were dispersed in water with a mass concentration of 50 μg/L. The NRs were characterized using transmission electron microscopy (TEM: FEI Titan Themis XFEG instrument) to investigate their structural and morphological properties, and the compositional analysis of the sample was performed using EDS inside the TEM. Additionally, UV-vis-NIR spectroscopy measurements were performed using a UV-vis-NIR spectrophotometer (Shimadzu UV-3600) to analyze the optical properties of the NRs.

**Optical Pump and XFEL Probe Setup.**

**X-ray Parameters.** The experiment was carried out using the Single Particles, Clusters, and Biomolecules Serial Femtosecond Crystallography (SPB/SFX) instrument[33] at the European X-ray Free-Electron Laser (EuXFEL) facility. The schematic of the experimental set-up is presented in Figure 2. The experiment utilized the EuXFEL's intra-train repetition rate of 564 kHz, delivering 202 pulses per train with an average pulse energy of approximately 0.8 mJ. The X-ray photon energy, beam size, and pulse width were 9.3 keV, 5 μm (H) × 40 μm (V) in FWHM, and ~25 fs, respectively.

**X-ray Detector.** TR-XRD data was collected using the Adaptive Gain Integrating Pixel Detector (AGIPD), with distance between the sample interaction region and the detector set to 122.2 mm.

**Sample Delivery.** We delivered the Au/Pd core-shell NRs using a 4.9 μm diameter water jet produced by a Gas Dynamic Virtual Nozzle (GDVN) with helium as the focusing gas, and the sample flow rate was about 20 μL/min, at a mass concentration of 20 μg/L.

**Optical Laser.** The NRs were excited by a circularly polarized 800 nm (wavelength) pump laser[52–54] with a pulse width stretched from an initial 50 fs to 890 fs using an SF57 glass rod. The laser wavelength was chosen to match well with the localized surface plasmon resonance (LSPR) of the major axis of the Au/Pd core-shell NRs, enabling efficient photothermal heating. A circularly polarized femtosecond laser was used to ensure isotropic excitation across the NR ensemble and suppress orientation-dependent absorption effects associated with the longitudinal plasmon resonance. The use of circular polarization was particularly important in our liquid jet geometry, where NRs exhibit partial alignment along the flow axis (see SI-2). Additionally, the stretched pulse duration of 890 fs was selected to reduce peak electric field strength, minimizing nonthermal reshaping pathways such as field enhancement or Coulomb-driven instabilities. These design choices ensured controlled thermal excitation and are further discussed in the Results and Discussions section in the context of lattice dynamics and energy absorption.

**Instrumental Broadening.**

Geometric contribution:

Detector pixel size 0.2 mm.

Sample to detector distance(z) 122 mm.

$$\Delta 2\theta = 0.2/122 \approx 0.0016 \, radians$$

Sample position uncertainty:

Any position on the diffraction ring from the direct beam position on the detector is x.

Sample was flowing in a water jet having a diameter of 0.005 mm.

So, the angular uncertainty Δ2θ due to positional uncertainty along the X-ray path

$$tan 2\theta = \frac{x}{z}$$

$$2\theta = tan^{-1}\left(\frac{x}{z}\right)$$

If the sample position moves along the beam direction Δz, then

$$2\theta = tan^{-1}\left(\frac{x}{z + \Delta z}\right)$$

$$\frac{d(2\theta)}{dz} = \frac{d}{dz}\left[tan^{-1}\left(\frac{x}{z + \Delta z}\right)\right] = -\frac{x}{x^2 + z^2}$$

So,

$$\Delta(2\theta) \approx \left|\frac{d(2\theta)}{dz}\right|.\Delta z = \frac{x}{x^2 + z^2}.\Delta z \approx 3.71 \times 10^{-5} \, rad$$

In our case, the resulting angular resolution is dominated by the detector pixel size.

During the measurement, the instrumental resolution contribution was 0.0016 rad.

**Time-Resolved X-ray Diffraction Scheme.** We employed a time-resolved X-ray diffraction setup in a pump-probe geometry to study the transient structural dynamics induced by a single optical excitation. To ensure single-pulse excitation, we carefully tuned the pulse-to-pulse separation, spot sizes of both the X-ray and optical laser beams, and the velocity of the liquid jet carrying the sample. The velocity of the water jet was measured to be approximately 40 m/s. Based on those, we calculated the nanoparticle's transit time across the ~40 μm diameter of the optical laser spot (Figure S10). The repetition rate of the optical pump laser was then adjusted such that the pulse-to-pulse separation exceeded this transit time, ensuring that each nanoparticle encountered only one optical pulse during its passage. Additionally, a cyclical laser on – laser off scheme was implemented to alternate between excitation and dark states. Diffraction data were collected at five different optical laser's fluences 2 mJ/cm², 10 mJ/cm², 20 mJ/cm², 100 mJ/cm², and 200 mJ/cm²—across fourteen-time delays (0 ps, 1 ps, 10 ps, 50 ps, 100 ps, 500 ps, 1 ns, 10 ns, 18.5 ns, 47 ns, 92.5 ns, 200 ns, 500 ns, and 1000 ns), allowing us to capture the ultrafast structural evolution from picosecond to microsecond timescales. The optical laser position was moved upward by 20 μm and 40 μm at delays of 500 ns and 1 μs, respectively, to compensate for the jet speed. The detailed calculation of the scheme is presented in Section SI-7.

**Data Analysis.** For each fluence–delay condition, ~15,000–20,000 individual diffraction patterns were collected. After applying filters for jet presence, detector integrity, signal strength, and correct ON/OFF tagging, ~200–500 high-quality patterns were retained per condition. Each usable frame corresponds to scattering from a small number of NRs (typically fewer than 5), given the low sample concentration (~20 μg/L), narrow 4.9 μm jet diameter, and focused XFEL beam profile. The high flow velocity (~40 m/s) ensured that each particle was excited only once, resulting in effectively single-event measurements with negligible re-illumination.

Data from individual patterns were integrated and averaged, and the error bars shown in Figures 3 and 4 represent the standard error of the mean across all retained high-quality data per condition. Sources of variability include detector shot noise, shot-to-shot misalignment within the jet, and intrinsic sample heterogeneity. The Au⟨111⟩ reflection was selected for primary analysis due to its high intensity, minimal overlap, and consistent behavior. Pd-related peaks were present but were weak and overlapped with Au reflections, and thus not used for quantitative analysis.

TR-XRD data were analyzed using custom Python scripts incorporating the PyFAI framework for azimuthal integration and detector calibration. Diffraction images collected by the AGIPD detector were corrected for dark current, geometry distortion, and beam center offset prior to radial integration. Bragg reflections were identified by comparing the experimental peak positions to standard patterns from the Inorganic Crystal Structure Database (ICSD), including entries for gold (ICSD 53763), palladium (ICSD 64920), palladium oxide (PdO; ICSD 77650), and the $Au_{1.51}Pd_{0.49}$ alloy (ICSD 197459). Peak fitting was performed using Gaussian functions to extract the peak position, peak width in FWHM, and integrated intensity of each peak (peak intensity × peak width). These parameters were used to calculate lattice constants, lattice strain, and crystallite sizes. Instrumental broadening (refer to Instrumental Broadening section) was subtracted before applying the Scherrer equation to determine crystallite size. Lattice temperature was derived from the temporal evolution of lattice expansion, and used to estimate the absorption cross-section via energy balance models. All laser-on datasets were compared to alternately collected laser-off datasets to isolate photoinduced structural effects. Additional calculations and uncertainty analysis can be found in the Supporting Information.

**Optical Pump and XFEL Probe Geometry.** For clarity, we define the experimental coordinate system as follows: the liquid jet propagates vertically along the negative y-axis, the XFEL beam is incident along the z-axis, and the optical laser is incident along the negative x-axis. The pump laser was circularly polarized in the y–z plane, i.e., perpendicular to the laser propagation direction. This geometry ensures that for nanorods aligned by shear flow along the jet axis (−y), the excitation field is effectively averaged over all orientations in the transverse plane. Any residual non-uniformity in excitation can only arise from deviations in the x–z plane, which are strongly reduced by the observed partial flow-induced alignment (~69% within a 15° cone, see SI-2.3). Together, this configuration minimizes orientation-dependent absorption and ensures that the ensemble-averaged TR-XRD signal reflects statistically representative heating and alloying dynamics.

## ASSOCIATED CONTENT

**Supporting Information**
The Supporting Information is available free of charge at
https:
SI-1. Size distributions and representative TEM images of core-shell Au/Pd (part number Pd-NR-250). Figures S1, S2, and S3. Table S1.
SI-2. Flow-induced alignment of nanorods inside the liquid jet. Figure S4.
SI-3. Calculation of optical absorption of core-shell nanorods from lattice expansion. Figures S5 and S6.
SI-4. Optical laser pump and the X-ray probe data. Figures S7 and S8.
SI-5. Compositional and morphological analysis of the alloy nanorods. Figure S9.
SI-6. Contribution of Debye-Waller factor to diffraction intensity.
SI-7. Calculation of the amount of sample in a liquid jet interacting with an optical laser. Figure S10.
SI-8. Comparative analysis of femtosecond laser-induced nanorod transformations. Tables S2 and, S3.

## AUTHOR INFORMATION

**Corresponding Authors**

**Abhisakh Sarma** – Single Particles, Clusters, and Biomolecules & Serial Femtosecond Crystallography (SPB/SFX), European X-ray Free-Electron Laser, Holzkoppel 4, Schenefeld, 22869, Germany; https://orcid.org/0000-0002-0785-8902 ; Email: abhisakh@gmail.com, abhisakh.sarma@xfel.eu
**Chan Kim** – Single Particles, Clusters, and Biomolecules & Serial Femtosecond Crystallography (SPB/SFX), European X-ray Free-Electron Laser, Holzkoppel 4, Schenefeld, 22869, Germany; https://orcid.org/0000-0003-4559-7982 ; Email: chan.kim@xfel.eu
**Tokushi Sato** – Single Particles, Clusters, and Biomolecules & Serial Femtosecond Crystallography (SPB/SFX), European X-ray Free-Electron Laser, Holzkoppel 4, Schenefeld, 22869, Germany; https://orcid.org/0000-0003-3155-3487 ; Email: tokushi.sato@xfel.eu

**Authors**

**Jayanath C. P. Koliyadu** − Single Particles, Clusters, and Biomolecules & Serial Femtosecond Crystallography (SPB/SFX), European X-ray Free-Electron Laser, Holzkoppel 4, Schenefeld, 22869, Germany; https://orcid.org/0000-0002-0245-3842
**Romain Letrun** – Single Particles, Clusters, and Biomolecules & Serial Femtosecond Crystallography (SPB/SFX), European X-ray Free-Electron Laser, Holzkoppel 4, Schenefeld, 22869, Germany; https://orcid.org/0000-0002-0569-5193
**Egor Sobolev** – Single Particles, Clusters, and Biomolecules & Serial Femtosecond Crystallography (SPB/SFX), European X-ray Free-Electron Laser, Holzkoppel 4, Schenefeld, 22869, Germany; https://orcid.org/0000-0003-2478-5685

**Trupthi Devaiah C** – Centre for Nano and Soft Matter Sciences (CeNS), Bengaluru 562162, India;
Present Address: Department of Chemistry, Ångström Laboratory, Uppsala University, Box 523, SE 75120 Uppsala, Sweden; https://orcid.org/0000-0002-5063-2992 ;
**Agnieszka Wrona** – Sample Environment & Characterization (SEC), European X-ray Free-Electron Laser, Holzkoppel 4, Schenefeld, 22869, Germany; https://orcid.org/0009-0004-2174-003X
**Katerina Doerner** – Sample Environment & Characterization (SEC), European X-ray Free-Electron Laser, Holzkoppel 4, Schenefeld, 22869, Germany; https://orcid.org/0000-0003-1072-5905
**Diogo V. M. Melo** – Single Particles, Clusters, and Biomolecules & Serial Femtosecond Crystallography (SPB/SFX), European X-ray Free-Electron Laser, Holzkoppel 4, Schenefeld, 22869, Germany; https://orcid.org/0000-0002-7417-3479
**Marco Kloos** – Sample Environment & Characterization (SEC), European X-ray Free-Electron Laser, Holzkoppel 4, Schenefeld, 22869, Germany; https://orcid.org/0000-0002-8745-9740
**Huijong Han** – Sample Environment & Characterization (SEC), European X-ray Free-Electron Laser, Holzkoppel 4, Schenefeld, 22869, Germany; https://orcid.org/0000-0002-1197-3014
**Marcin Sikorski** – Single Particles, Clusters, and Biomolecules & Serial Femtosecond Crystallography (SPB/SFX), European X-ray Free-Electron Laser, Holzkoppel 4, Schenefeld, 22869, Germany; https://orcid.org/0000-0001-7930-7790
**Konstantin Kharitonov** – Single Particles, Clusters, and Biomolecules & Serial Femtosecond Crystallography (SPB/SFX), European X-ray Free-Electron Laser, Holzkoppel 4, Schenefeld, 22869, Germany; https://orcid.org/0000-0002-4707-1047
**Juncheng E** – Single Particles, Clusters, and Biomolecules & Serial Femtosecond Crystallography (SPB/SFX), European X-ray Free-Electron Laser, Holzkoppel 4, Schenefeld, 22869, Germany; https://orcid.org/0000-0001-6061-5734
**Joana Valerio** – Sample Environment & Characterization (SEC), European X-ray Free-Electron Laser, Holzkoppel 4, Schenefeld, 22869, Germany; https://orcid.org/0000-0001-5931-0925
**Pralay K. Santra** – Centre for Nano and Soft Matter Sciences (CeNS), Bengaluru 562162, India; https://orcid.org/0000-0002-9358-5835 ;
**Erik M. J. Johansson** - Department of Chemistry, Ångström Laboratory, Uppsala University, Box 523, SE 75120 Uppsala, Sweden; https://orcid.org/0000-0001-9358-8277
**Richard Bean** – Single Particles, Clusters, and Biomolecules & Serial Femtosecond Crystallography (SPB/SFX), European X-ray Free-Electron Laser, Holzkoppel 4, Schenefeld, 22869, Germany; https://orcid.org/0000-0001-8151-7439

**Notes**
The authors declare no competing financial interest.

### ACKNOWLEDGMENTS

We acknowledge European XFEL GmbH in Schenefeld, Germany, for provision of X-ray free-electron laser beamtime at the SPB/SFX instrument (internal beamtime proposal numbers 4856 and 5996), SEC group for providing sample delivery and, we thank the instrument group and facility staff for their great assistance. We acknowledge the use of the TEM facility at Myfab Uppsala, supported by the Swedish Research Council (2020-00207) as part of the national research infrastructure. We acknowledge Professor G. U. Kulkarni, Director (I/c), Centre for Nano and Soft Matter Sciences (CeNS), for his valuable suggestions regarding the interpretation of the electron microscopy data.

Electronic supporting information for:

# Time Resolved Study of Laser-Induced Ultrafast Alloying Processes in Au/Pd Core–Shell Nanorods

Abhisakh Sarma*1, Jayanath C. P. Koliyadu1, Romain Letrun1, Egor Sobolev1, Trupthi Devaiah C2, †, Agnieszka Wrona1, Katerina Doerner1, Diogo V. M. Melo1, Marco Kloos1, Huijong Han1, Marcin Sikorski1, Konstantin Kharitonov1, Juncheng E1, Joana Valerio1, Pralay K. Santra2, Erik M. J. Johansson3, Richard Bean1, Chan Kim*1, Tokushi Sato*1

1 European Free-Electron Laser, Holzkoppel 4, Schenefeld, 22869, Germany
2 Centre for Nano and Soft Matter Sciences (CeNS), Bengaluru 562162, India
†, 3 Department of Chemistry, Ångström Laboratory, Uppsala University, Box 523, SE 75120 Uppsala, Sweden

**Contents:**

**SI-1. Size distributions of Au/Pd core-shell nanorods and representative TEM images. Figures S1, S2, and S3.**

**SI-2. Flow-induced alignment of nanorods inside the liquid jet. Figure S4 and Table S1.**

**SI-3. Calculation of optical absorption of Au/Pd core-shell nanorods based on lattice expansion. Figures S5 and S6.**

**SI-4. Time-resolved X-ray diffraction data (optical laser pump and X-ray probe). Figures S7 and S8.**

**SI-5. Compositional and morphological analysis of the Au/Pd alloy nanorods. Figure S9.**

**SI-6. Contribution of Debye-Waller factor to diffraction intensity.**

**SI-7. The amount of sample in a liquid jet interacting with an optical laser pulse. Figure S10.**

**SI-8. Comparative analysis of femtosecond laser-induced nanorod transformations. Tables S2 and S3.**

† Present address:

**SI-1. Size distributions of Au/Pd core-shell nanorods and representative TEM images:**

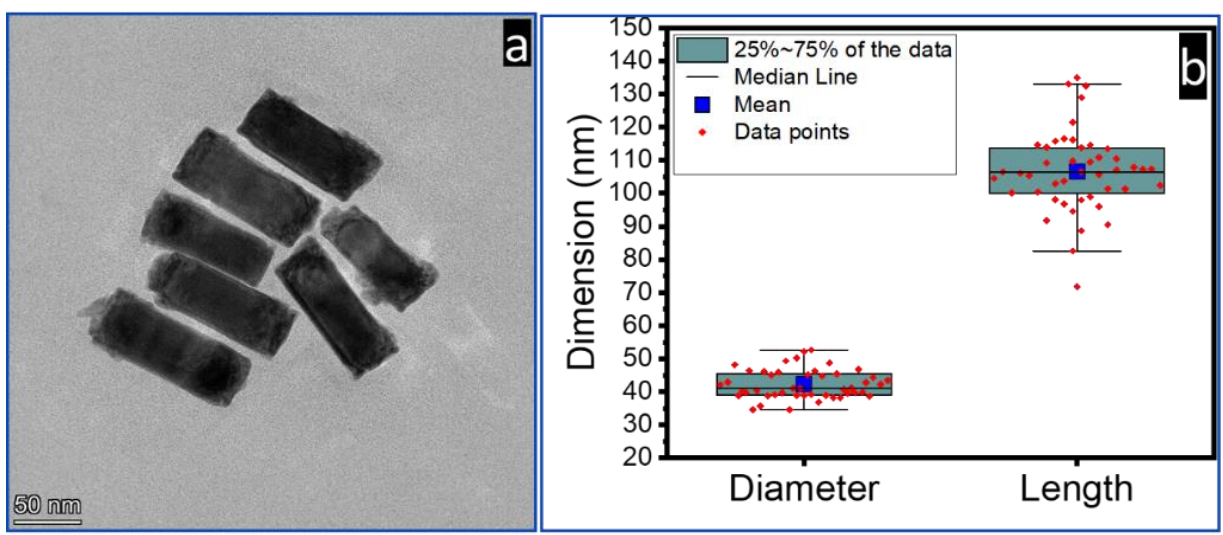

*Figure S1: (a) Transmission electron microscopy (TEM) image of Au/Pd core-shell nanorods, showing their morphology and size distribution. (b) Statistical analysis of the measured dimensions, represented as a box plot displaying the distribution of data points, mean, and median values.*

To estimate the size distribution, we analyzed several tens of low-resolution TEM micrographs, with the analysis performed using ImageJ software, as shown in Figure S1. Based on the analysis shown in Figure S1, the mean and standard deviation of the length and the diameter distribution (in nm) are as follows:

Length: 106 nm ± 12 nm (mean ± standard deviation).
Diameter: 44 nm ± 4 nm (mean ± standard deviation).

| Length: | Diameter: |
|---|---|
| Mean: 106 nm | Mean: 44 nm |
| Standard Deviation: 12 nm | Standard Deviation: 4 nm |
| Minimum: 72 nm | Minimum: 35 nm |
| Maximum: 132 nm | Maximum: 53 nm |
| 25th Percentile: 100 nm | 25th Percentile: 40 nm |
| Median (50th Percentile): 106 nm | Median (50th Percentile): 41 nm |
| 75th Percentile: 112 nm | 75th Percentile: 46 nm |

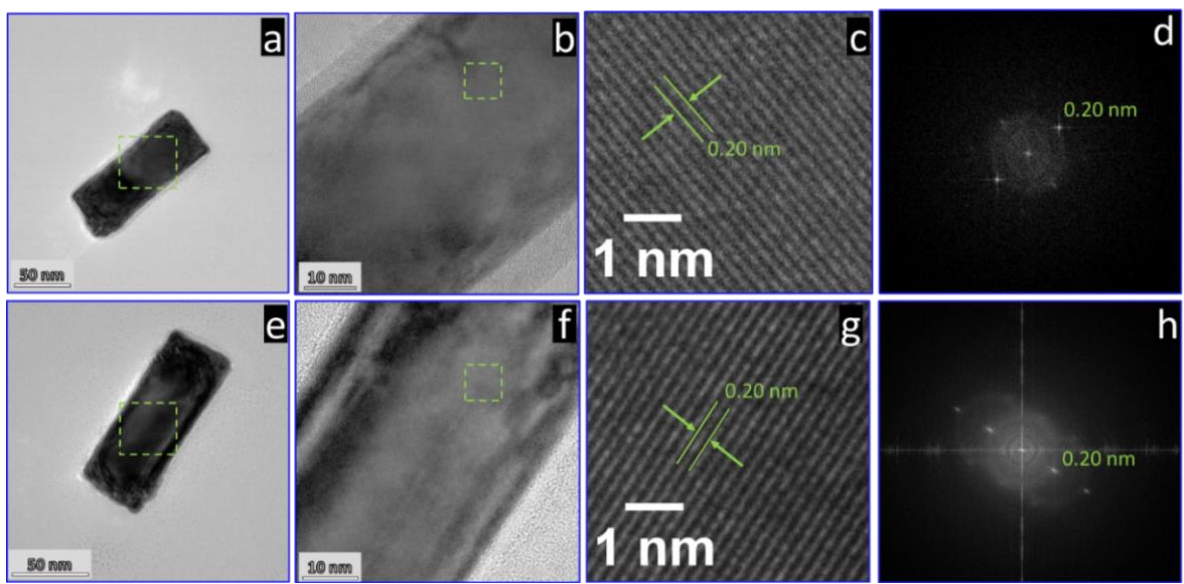

*Figure S2: (a) and (e) TEM images of Au/Pd core-shell nanorod showing the overall morphology. (b) and (f) HRTEM images of the selected regions (dashed rectangles in (a) and (e)). (c) The magnified image of the selected HRTEM image marked as dashed rectangle in (b), highlighting the (200) plane oriented along the length of the nanorod. (d) Fast Fourier Transform pattern of the HRTEM image (c), indicating a lattice spacing of 0.20 nm, corresponding to the (200) plane of Au. (g) The magnified image of the selected HRTEM image marked as dashed rectangle in (f), highlighting the (200) plane oriented across the diameter of the nanorod. (h) Fast Fourier Transform pattern of the HRTEM image (g).*

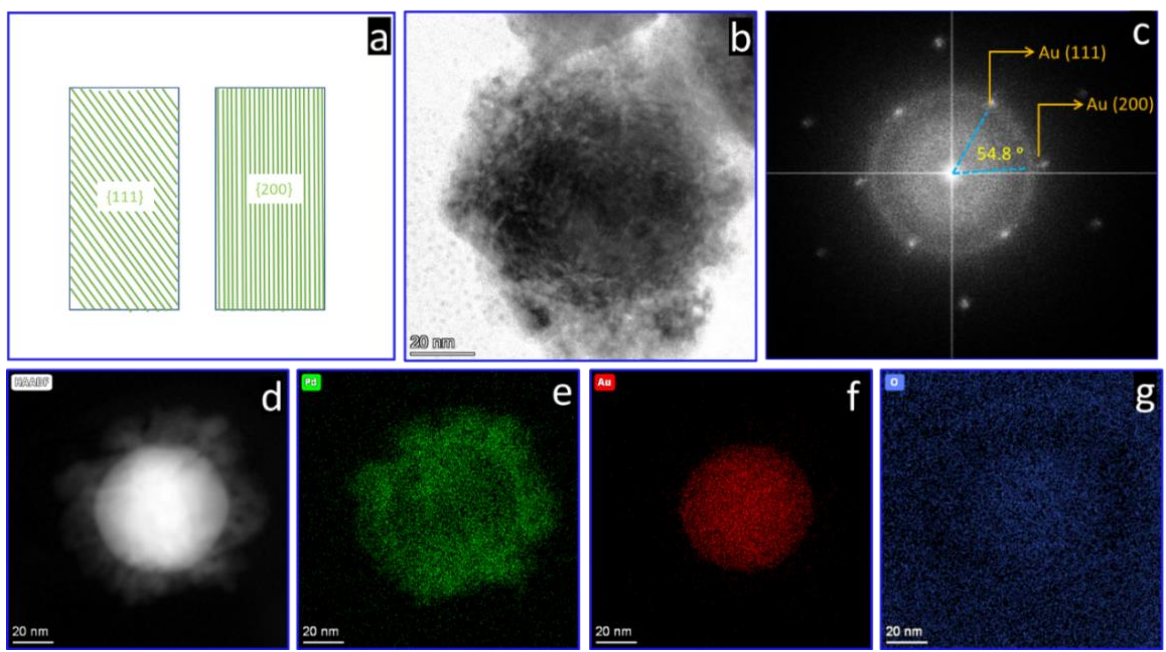

*Figure S3: (a) Schematic illustration of the {111} and {200} crystal facets of the gold core of the Au/Pd core-shell nanorod. (b) TEM image of the top view of the nanorod. (c) Fast Fourier Transform pattern of the top view of the nanorod shown in (b), confirming the presence of Au (111) and Au (200) planes, with an interplanar angle of 54.8°. (d) High-angle annular dark-field scanning transmission electron microscopy (HAADF-STEM) image of the top view of the nanorod. (e–g) Energy-dispersive X-ray spectroscopy (EDS) elemental mapping images showing the distribution of (e) Pd (green), (f) Au (red), and (g) O (blue), indicating the composition of the sample.*

## SI-2. Flow-induced alignment of nanorods inside the liquid jet:

### SI-2.1. Inflow nanorods' characterization with X-ray diffraction:

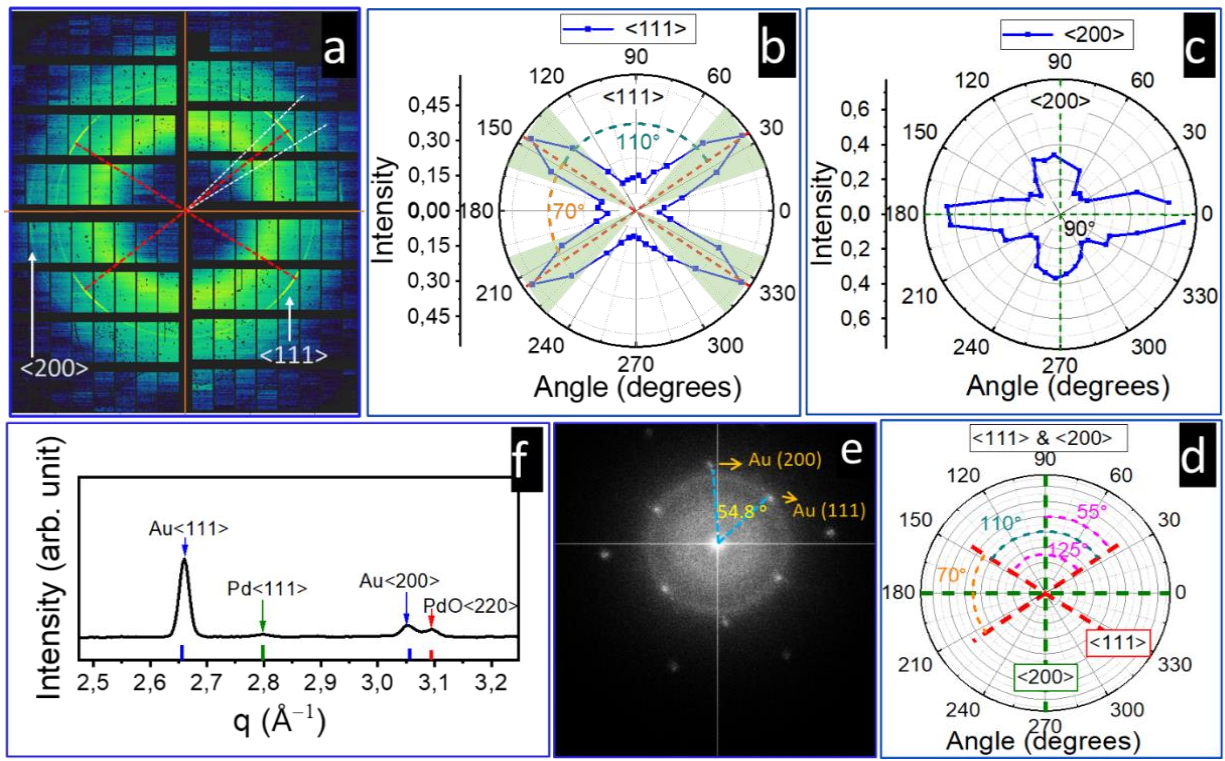

*Figure S4: (a) Averaged 2D X-ray diffraction (XRD) pattern of pristine Au/Pd nanorods measured during their flow in a water jet, highlighting the preferential orientation of the {111} and {200} reflections (averaged over around 300 frames). The red and the white dotted lines correspond to the maximum intensity position and 10° azimuthal slice, respectively. (b), (c) Polar plots of the intensity (b) Au⟨111⟩ and (c) Au⟨200⟩ Bragg reflections, showing angular dependencies on the detector plane. (d) Comparative orientation analysis of ⟨111⟩ and ⟨200⟩ reflections, displaying characteristic angular separations. (e) Fast Fourier Transform pattern of the nanorods (Figure S3b), with indexed Au(111) and Au(200) planes, and a measured interplanar angle of 54.8°. (f) One-dimensional line profile obtained from the averaged 2D XRD data shown in (a).*

In situ XRD measurements were conducted to investigate the inflow orientational dependence of the nanorods (NRs), and for material identification. The average 2D diffraction data, collected on the AGIPD detector, is presented in Figure S4(a). A representative one-dimensional integrated diffraction pattern is shown in Figure S4(f), featuring prominent peaks corresponding to Au ($q \approx 2.66$ Å$^{-1}$, (111)), Au ($q \approx 3.05$ Å$^{-1}$, (200)), Pd ($q \approx 2.79$ Å$^{-1}$, (111)), and a weaker peak attributed to PdO ($q \approx 3.09$ Å$^{-1}$, (220)), which further confirmed the presence of oxygen in the palladium-rich shell of the nanorod, as shown in Figure S3(g). For randomly oriented nanorods in the water jet, we expected to observe a circular diffraction ring with uniform intensity around its perimeter. However, we observed textured partial rings corresponding to the <111> and <200> planes of gold, suggesting a preferential alignment of the nanorods during the flow, as shown in Figure S4(a).

Figures S4(b) and S4(c) present the azimuthal angle dependence peak area plots for both the <111> and <200> planes. To generate these plots, we performed a 1D integration of the averaged 2D data over 360° in 10° angular intervals (36 steps). The area under the <111> and <200> reflections for each 10° interval was plotted against the azimuthal angle in polar coordinates, with the polar angle representing the azimuthal angle on the detector plane. The polar plot of the <111> reflection shows four lobes with nearly equal intensity, separated by mutual angles of 70° or 110°. In contrast, the polar plot of the <200> reflection also displays four lobes, but the horizontal lobes are more intense than the vertical ones. The mutual angle between the lobes for the <200> planes is 90°. Figure S4(d) shows the angular locations of the <111> and <200> planes, marked with red and green dotted lines, respectively. From this figure, we can observe that the mutual angle between the lobes of the <111> and <200> planes is either 55° or 125°.

The angular relationship between the <111> and <200> planes, as derived from the polar plot, matches well with the angular relationship observed in the Fast Fourier transform of the crystallographic planes from the HRTEM image of the nanorods, as shown in Figures S3(c) and S4(e). In a face-centered cubic (FCC) unit cell, the mutual angles between the <111> planes are 70° and 109.47°, while the mutual angle between the four-fold symmetric <200> planes is 90°.

The angle between the <111> and <200> planes is either 54.74° or 125.26°. Therefore, the angular relationships observed in the XRD data are consistent with those found between different planes in the FCC unit cell structure. By comparing the angular relationships of the lobes in the averaged 2D XRD data with the HRTEM images, we conclude that the nanorods were primarily aligned parallel to the jet direction during the jetting process triggered by flow-induced alignment due to the anisotropic shape of the nanorods. A calculation of the flow-induced alignment of the nanorods is presented below.

**SI-2.2. Calculations for the probability of the inflow alignment of nanorods:**

Nanorod parameters:
Length of nanorods (L): 106 nm = 106×10⁻⁹ m
Diameter of nanorods (d): 44 nm = 44×10⁻⁹ m
Mass-concentration of nanorods: 20 µg/L = 20×10⁻⁶ g/L = 20×10⁻⁹ kg/L
Density of gold ($\rho_{Au}$): 19300 kg/m³
Water density ($\rho_{water}$): 1000 kg/m³

$$V_{nanorod} = \pi \left(\frac{d}{2}\right)^2 L = 1.61 \times 10^{-22} \text{ m}^3$$

$$m \text{ (nanorod)} = 19{,}300 \times 1.61 \times 10^{-22} \approx 3.11 \times 10^{-18} \text{ kg}$$

Flow parameters:

Jet velocity (v): 40 m/s
Water jet diameter ($D_{jet}$): 4.9 µm = 4.9 × 10⁻⁶ m

Reynolds number calculation[1,2]:
The Reynolds number (Re) helps to determine whether the flow is laminar or turbulent and can indicate how nanorods interact with the flow. For the water jet, we calculate the Reynolds number:

$$Re = \frac{\rho_{water} \upsilon D_{jet}}{\mu_{water}}$$

$$Re = \frac{1000 \times 40 \times 4.9 \times 10^{-6}}{\mu_{water}} = 196$$

Dynamic Viscosity of Water ($\mu_{water}$)(approximate at 20° C is 1.002×10⁻³ kg/(m.s)

Brownian motion consideration:
The nanorods will also undergo Brownian motion due to thermal energy. The Brownian motion can be quantified using the diffusion coefficient for a particle:

$$D = \frac{k_B T}{6\pi\eta r}$$

where:
$k_B$ is Boltzmann's constant (1.38×10⁻²³ J/K),
T is the temperature in Kelvin (assume T=298 K).
Substituting the values:
D = 1.74 x 10⁻¹³ m²/s
This is the diffusion coefficient for a single nanorod.
Peclet number (advection vs. diffusion)[3]
Next, we calculate the Peclet number (Pe), which compares the effects of advection (flow) to diffusion:

$$Pe = \frac{106 \times 10^{-9} \times 40}{1.73 \times 10^{-13}} \approx 2.4 \times 10^7$$

where:
L = 106 nm = 106×10⁻⁹ m (length of the nanorod),
v = 40 m/s (velocity of the fluid),

D = 1.74×10$^{-13}$ m$^2$/s (diffusion coefficient for the nanorods, calculated earlier).

Flow-induced ordering probability:

A high Peclet number (Pe ≈ 2.4 × 10$^7$) suggests that advection (the flow velocity) dominates over diffusion. This means the nanorods will be strongly influenced by the flow and will likely align with the direction of the flow.

Given that the Reynolds number is 196 (indicating laminar flow), the nanorods will experience steady, predictable flow forces, and because advection dominates, they are very likely to align with the flow direction.

Conclusion:

The flow-induced ordering probability of the Au/Pd core-shell nanorods is extremely high, given the dominant advection effect (high Peclet number) and laminar flow conditions (Reynolds number of 196). The nanorods will align along the direction of the flow with a very high probability.

**SI-2.3. Quantification of flow-induced nanorod alignment:**

To quantify the degree of nanorod alignment within the liquid jet, we analyzed the azimuthal intensity profile of the Au⟨111⟩ reflection (Figure b). Flow-induced ordering aligns the nanorods' long axis along the jet direction, resulting in four distinct, sharp diffraction lobes in the azimuthal distribution, centered around ≈ 35.5°, 145.5°, 215.5°, and 325.5°.

We defined the "aligned" population as the integrated intensity within ±15° cones around these four main lobe centers, as visually indicated by the transparent green cones in Figure S4(b). This angle was chosen to balance the need to capture the full width of the narrow alignment peak against the need to exclude the underlying background intensity and contributions from misaligned nanorods. While our data resolution is defined by a 10° azimuthal integration step, the ±15° cone ensures a robust quantification by encompassing the central bin and the immediate neighboring bins (a total span of 30°), thereby accounting for the true angular spread of the flow-aligned population.

| Cone center (°) | Angles included (°) | Intensities | Cone sum |
|---|---|---|---|
| 35.5 | 25.5, 35.5, 45.5 | 0.377, 0.522, 0.403 | 1.302 |
| 145.5 | 135.5, 145.5, 155.5 | 0.435, 0.519, 0.316 | 1.270 |
| 215.5 | 205.5, 215.5, 225.5 | 0.491, 0.52, 0.364 | 1.375 |
| 325.2 | 315.5, 325.5, 330.5 | 0.355, 0.53, 0.385 | 1.27 |
| All angles (0–360°) | — | Sum of all intensities = 7.57 | — |

*Table S1: Azimuthally integrated Au⟨111⟩ diffraction intensities used to quantify flow-induced nanorod alignment. Intensities within ±15°(marked as green cone in Figure S4b) of the four main lobe centers (35.5°, 145.5°, 215.5°, 325.5°) were summed to define the aligned fraction. The final row provides the total integrated intensity over the full 0–360° range.*

**Methodology:**

The integrated azimuthal intensity I(ϕ) over 0 – 360° represents the contribution of all nanorods to diffraction.

Flow-induced ordering aligns the nanorods' long axis along the jet direction. Because the nanorods grow along the crystallographic <111> axis, this ordering produces four distinct lobes in the azimuthal distribution, separated by ~110° and 70°.

We defined the "aligned" population as the integrated intensity within ± 15° cones around the four main lobe centers (≈ 35.5°, 145.5°, 215.5°, 325.5°).

The fraction of aligned nanorods was calculated as:

Alignment Fraction = (∑Itotal / ∑Icones) × 100%

**Results:**

Total integrated intensity over 360°: 7.57 (arbitrary units).

Aligned intensity (sum over four cones): 5.21.

Alignment fraction: 5.21/7.57×100% ≈ 68.8 %

**Conclusion:**

Approximately 69 % of the nanorods were aligned within ± 15° of the jet axis, confirming a strong degree of preferential orientation induced by the liquid jet. The remaining ∼ 31 % are distributed outside the cones, consistent with partial, rather than perfect, alignment.

**SI-3. Calculation of optical absorption of Au/Pd core-shell nanorods based on lattice expansion:**

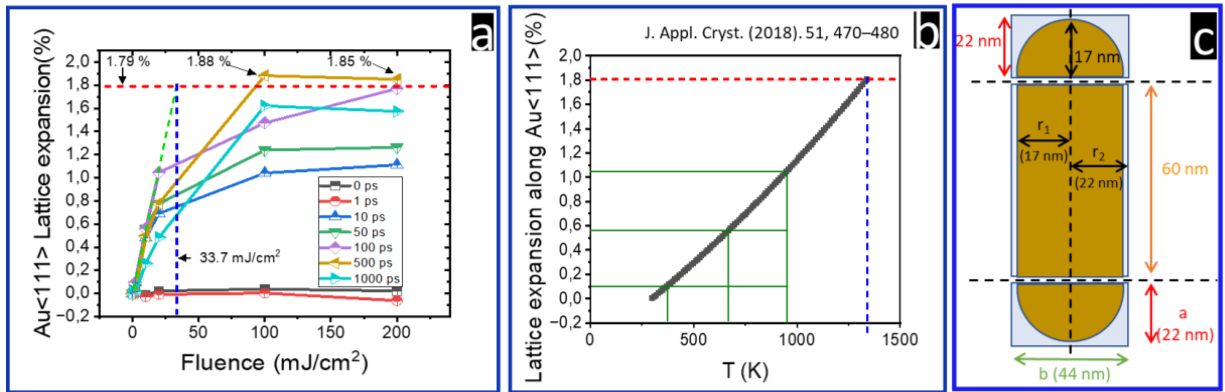

*Figure S5: (a) Lattice expansion of Au<111> as a function of laser fluence for different time delays (0 ps to 1000 ps). The dashed horizontal red line represents the maximum expansion limit of gold before melting, the green dashed line with a slope represents the linear fitted extrapolated line of the 100 ps data, while the dashed vertical blue line indicates the intersection point between the green and red dashed lines. (b) Theoretical lattice expansion along Au<111> as a function of temperature, adapted from J. Appl. Cryst. (2018), 51, 470–480. (c) Schematic representation of the nanostructure used in the study, showing key geometric dimensions.*

**SI-3.1. Thermal expansion and lattice temperature calculation:**

The temperature-dependent volume expansion of gold is,

$$V(T) = V_{T_r} \exp\left[\int_{T_r}^{T_f} \alpha(T)dT\right]$$

Here[4,5], $T_r$ and $T_f$ correspond to the initial and final temperatures, respectively. $\alpha(T)$ is the temperature-dependent volume expansion coefficient which can be expressed as

$$\alpha(T) = a_0 + a_1 T$$

The used values[6,7] for $a_0$ and $a_1$ are 3.62 (2) × 10$^{-5}$ K$^{-1}$ and 1.88 (3) × 10$^{-8}$ K$^{-2}$, respectively.

Using the above-mentioned information, we generated the graph presented in Figure S5(b). By using the value of thermal expansion extracted from Figure S5(a), we extracted the values of lattice temperature, which correspond to specific values of used laser fluences. In our case, the measured lattice expansions from the laser fluences of 2 mJ/cm², 10 mJ/cm², and 20 mJ/cm² are 0.0954 %, 0.56045 % and 1.0459 %, respectively. The derived lattice temperature at fluence values of 2 mJ/cm², 10 mJ/cm², and 20 mJ/cm² are 370 K, 670 K, and 950 K, respectively.

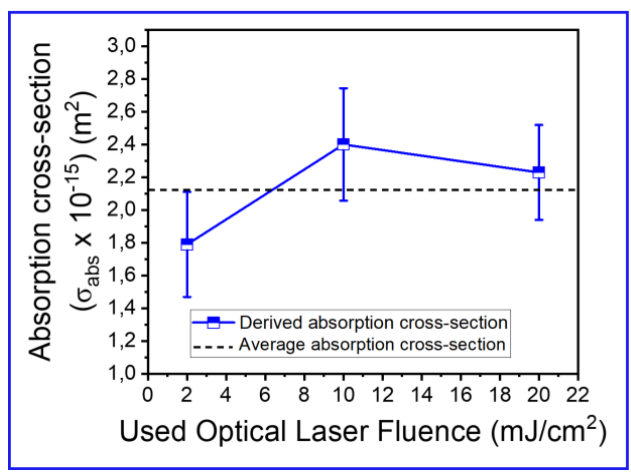

*Figure S6: Plot of the calculated absorption cross-section derived from measured lattice expansion. The 10 % size distribution (both along the length and across the diameter) which resulted in the uncertainties for the surface area change (± 15 %) which we considered as error bar.*

The optical absorption cross-section was estimated by back-calculating the absorbed energy per nanorod using the measured lattice expansion, thermal expansion coefficients, and known heat capacities. This yielded a value of approximately $2.13 \times 10^{-15}$ m² under XFEL-matched fluence and pulse conditions. Although we recorded UV–vis extinction spectra for the nanorods (see Figure 1h), these were not used for absolute calibration due to uncertainties in the optical path length, scattering, and particle concentration in the cuvette.

As discussed by González-Rubio *et al.*[8], determining absorbed energy per particle in solution-based systems is complicated by random orientation and aspect ratio heterogeneity, which lead to variability in the optical absorption efficiency. These limitations precluded accurate UV–vis based energy quantification in their single-pulse reshaping experiments. Our TR-XRD approach overcomes these issues by providing a direct structural readout of energy deposition. Partial nanorod alignment from flow (see SI-2), use of circularly polarized pump laser, and analysis of the Au⟨111⟩ peak allow us to recover a physically meaningful absorption estimate. This value agrees in order of magnitude with previously reported absorption cross-sections for pure Au nanorods of similar dimensions[9], supporting the reliability of our thermal modeling.

**SI-3.2. Photon absorption calculation from the lattice temperature:**

The photon absorption cross-section ($\sigma_{abs}$), laser fluence (F), specific heat (Cp), Temperature change ($\Delta T$), and mass of the particle (m) are related to each other with the relation

$$m\ C_p\ \Delta T = \sigma_{abs}\ F$$

Using values of lattice temperatures attained for different laser fluence values mentioned in the last section and the Cp values from the published literature[10], we can calculate the photon absorption cross-section. The calculated values of the absorption factor for different laser fluence values are presented in Figure S6.

**SI-3.3. Required heat for the melting of bulk gold:**

The melting temperature of bulk gold is 1337.33 K

The specific heat of gold in the temperature range between 298.15 K to 1337.33 K is written as[7]

$$C_p^0 \left(\frac{J}{mol\ K}\right) = 25.7663 - 1.715184 \times 10^{-3}\ T + 4.95270 \times 10^{-6}\ T^2 - \frac{37708.8}{T^2}$$

The amount of required enthalpy to raise the temperature of gold to the melting point is

$$\text{Enthalpy} = V(T) = V_{Tr} \exp\left[\int_{298.15}^{1337.33} C_p^0\ dT\right] = 29.124\ kJ/mol$$

The required free energy of only for melting is 12.72 kJ/mol. Hence, the energy needed to completely melt gold is (29.124 + 12.72) kJ/mol = 41.84 kJ/mol.

**SI-3.4. Energy gained by the atoms of the nanorods:**

Our estimated effective absorption cross-section (from lattice-expansion back-calculation, presented in SI-3): $\sigma_{abs} \approx 2.13 \times 10^{-15}\ m^2$
(We state this is an effective cross-section for our experimental geometry (see SI-3.)

Estimated fluence threshold:

F = 48 mJ/cm² = 48 ×10−3 J/cm² = 480 J/m²

Nanorod volumes (from SI-5):

Gold core volume, $V_{Au}$ = 99077 nm³

Palladium shell volume, $V_{Pd}$ = 77338 nm³

Bulk densities and molar masses:

ρ(Au)=19300 kg·m⁻³, M(Au)=196.97 g·mol⁻¹,

ρ(Pd)=12023 kg·m⁻³, M(Pd)=106.42 g·mol⁻¹.

Number of atoms per nanorod (approximate):

Volume conversion to m³: $V_{Au} = 9.9077 \times 10^{-23}\ m^3$; $V_{Pd} = 7.7338 \times 10^{-23}\ m^3$

Masses: $m_{Au} = \rho_{Au} \times V_{Au} \approx 1.912 \times 10^{-18}\ kg$; $m_{Pd} \approx 9.298 \times 10^{-19}\ kg$.

Number of atoms (mass → moles → atoms):
$N_{Au} \approx 5.85 \times 10^6$ atoms,
$N_{Pd} \approx 5.26 \times 10^6$ atoms.
$N_{total} \approx 1.11 \times 10^7$ atoms per nanorod.

Absorbed energy per nanorod (upper bound, η = 1, for full conversion):

$E_{abs} = F \times \sigma_{abs} = 480$ J/m² $\times 2.13 \times 10^{-15}$ m² $\approx 1.02 \times 10^{-12}$ J.

Absorbed energy per atom (upper bound):

$$E_{atom} = \frac{E_{abs}}{N_{total}} \approx \frac{1.02 \times 10^{-12} J}{1.11 \times 10^7} = 9.2 \times 10^{-20} \text{ J/atom} = 0.57 \text{ eV/atom}$$

[1 J → 6.242 x $10^{18}$ eV]

Comparison to melting enthalpy per atom (SI-3):

Enthalpy to raise Au to melt + latent heat (SI-3): ≈ 41.84 kJ·mol$^{-1}$ → per-atom ≈ $6.95 \times 10^{-20}$ J/atom ≈ 0.434 eV/atom.

Our upper-bound absorbed energy ≈ 0.57 eV/atom at 48 mJ/cm² is comparable to and slightly above the energy required to melt Au. This is consistent with the interpretation that transient Au melting may accompany lattice-level Au–Pd interdiffusion under our experimental conditions. (subject to the caveats below).

**Important caveats and uncertainties:**

Upper-bound assumption on conversion efficiency (η)

We assumed all absorbed photon energy is converted into lattice heating. In reality, some energy is lost via radiative scattering, hot-electron escape, or other nonthermal channels. Thus 0.57 eV/atom is an upper bound; the real deposited lattice energy will be somewhat lower.

**Absorption cross-section uncertainty:**

The effective $\sigma_{abs}$ is itself derived from lattice-expansion modeling and has experimental/model uncertainties (size polydispersity, partial alignment, beam overlap). SI-3 notes ~ ± 15% uncertainty from size distribution and other factors — propagate that to the per-atom number.

**Non-uniform energy deposition inside the particle.**

Local heating (e.g., tips, shell vs core) will be nonuniform; the per-atom number above is an average. Local superheating at interfaces or hot-spots can exceed the mean and drive interdiffusion.

**Heat flow & phase-change dynamics.**

Not all absorbed energy goes immediately into melting; some will be used to raise temperature, some will drive latent heat, and ultra-fast nonequilibrium processes can transiently store or re-distribute energy.

**Pd involvement.**

Melting/alloying requires consideration of both Au and Pd thermodynamics we compared to Au enthalpy because Au core melting is the critical event for interdiffusion, but shell properties and mixing enthalpies matter for kinetics.

**SI-4. Time-resolved X-ray diffraction data (optical laser pump X-ray probe):**

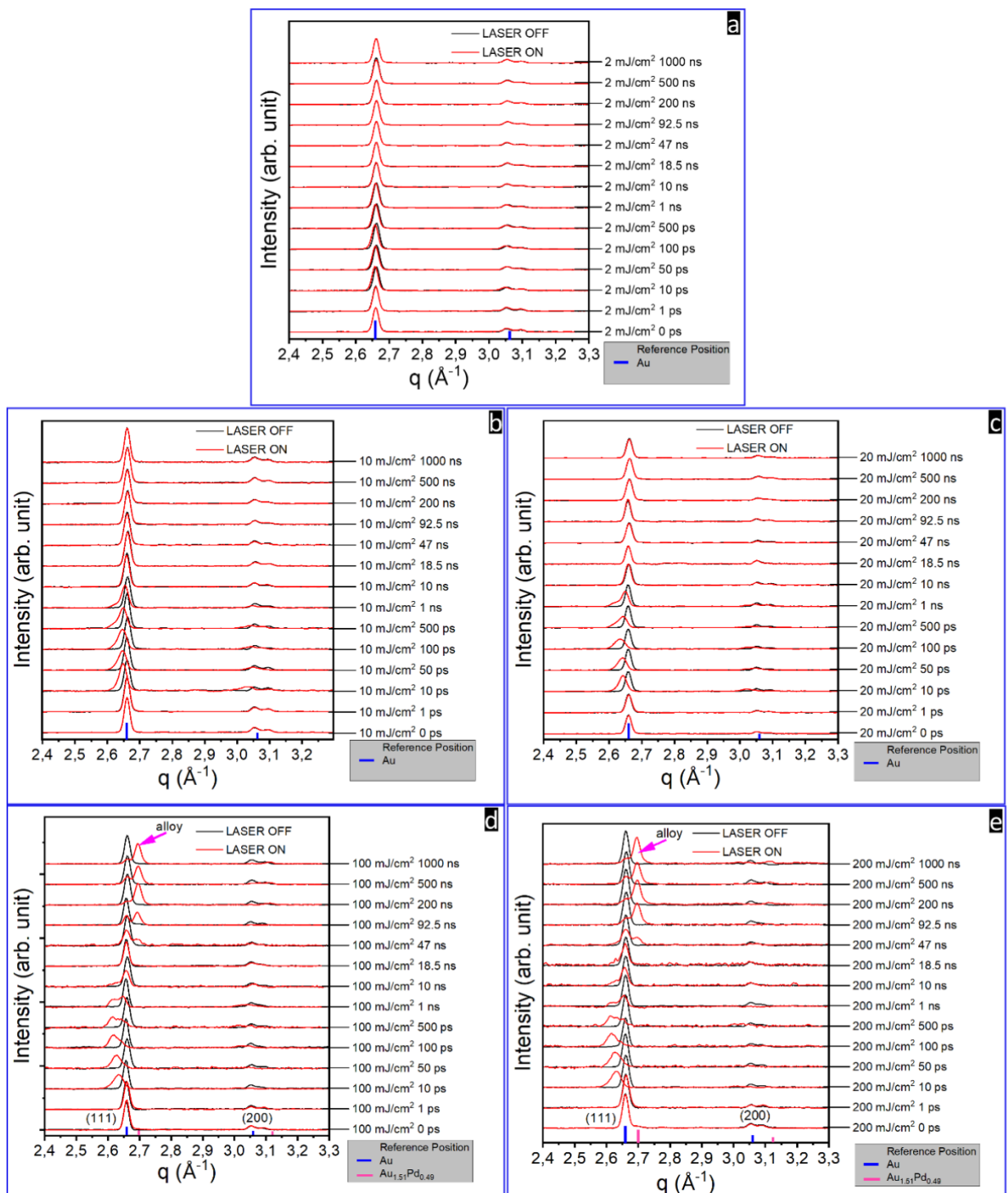

*Figure S7: Time-resolved X-ray diffraction data showing the evolution of diffraction peaks under 800 nm (wavelength) and 890 fs (pulse duration) laser excitation. The black curves represent the diffraction pattern before laser excitation (LASER OFF), while the red curves correspond to the laser-excited state (LASER ON) at various time delays. The measurements were conducted at a fluence of (a) 2 mJ/cm², (b) 10 mJ/cm², (c) 20 mJ/cm², (d) 100 mJ/cm², and (e) 200 mJ/cm², with varying time delays ranging from 0 ps to 1000 ns. The blue marker indicates the reference lattice constant of gold. [gold (ICSD 53763), and the $Au_{1.51}Pd_{0.49}$ alloy (ICSD 197459)].*

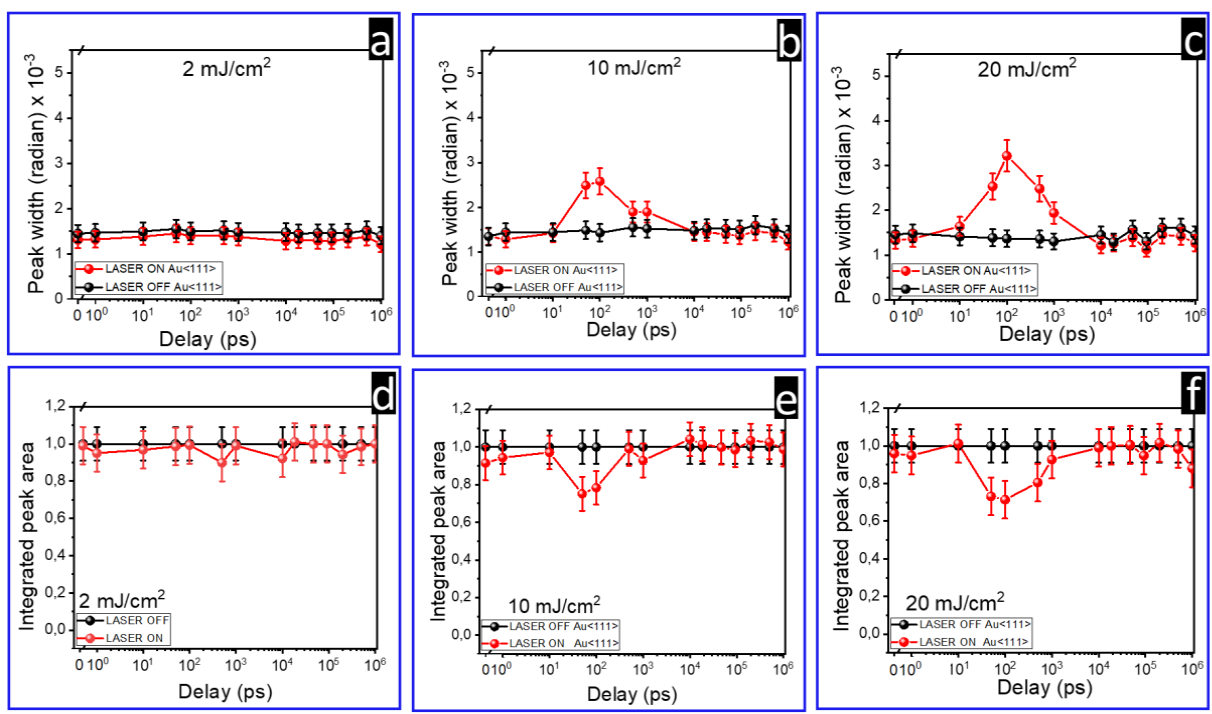

*Figure S8: Time-resolved X-ray diffraction analysis of Au<111> under different laser fluences. (a–c) Evolution of the peak width as a function of pump-probe delay for the laser fluences of 2 mJ/cm², 10 mJ/cm², and 20 mJ/cm², respectively. (d–f) Integrated peak area for the same fluences. The red and black data points correspond to measurements with and without laser excitation, respectively. A transient increase in peak width and a corresponding decrease in peak area are observed for higher fluences, indicating the occurrence of melting of gold. Error bars represent standard deviations. We applied a scale break at 0.5 ps on the x-axis. The section before the break is linear, while the section after the break follows a logarithmic scale.*

**SI-5. Compositional and morphology analysis of the Au/Pd alloy nanorods:**

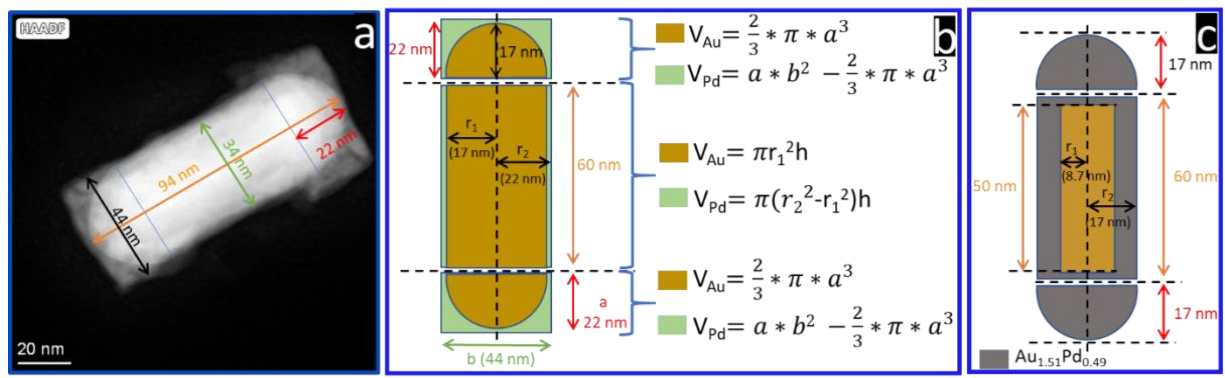

*Figure S9: (a) High-angle annular dark field (HAADF) scanning transmission electron microscopy (STEM) image of the nanostructure, showing its dimensions. (b) Schematic representation of the nanostructure with key geometric parameters highlighted. (c) Schematic illustrating the core-shell geometry of the alloy nanorod (final product) with relevant size annotations (200 mJ/cm² at 1 μs time delay).*

**Volume and atomic ratios in pristine Au/Pd core-shell nanorods:**

The volume of the pristine gold rod ($V_{Au}$) is 99077 nm³, and the volume of the pristine palladium rod ($V_{Pd}$) is 77338 nm³. The lattice constant for the face-centered cubic (FCC) gold ($a_{Au}$) is 0.409 nm, while the lattice constant for the FCC palladium ($a_{Pd}$) is 0.3098 nm. The atomic ratio between gold and palladium in the pristine core-shell nanorod is given by:

$$\text{Atomic ratio} = \frac{V_{Au}}{V_{Pd}} \times \left(\frac{\text{Volume of Pd unit cell}}{\text{Volume of Au unit cell}}\right).$$

Substituting the known values:

$$\text{Atomic ratio} = \frac{99077}{77338} \times \left(\frac{0.3098}{0.409}\right)^3 = 0.56$$

The remaining volume of gold ($V_{Au, \text{remaining}}$) is calculated as:

$$V_{Au, \text{remaining}} = 99077 \times 0.16 = 15852 \text{ nm}^3 \text{ [at laser fluence 100 mJ/cm}^2\text{]}$$

$$V_{Au, \text{remaining}} = 99077 \times 0.12 = 11889 \text{ nm}^3 \text{ [at laser fluence 200 mJ/cm}^2\text{]}$$

Diameter of remaining gold core calculation:

$$\pi \times \left(\frac{d}{2}\right)^2 \times L = V_{Au, \text{remaining}}$$

In case of the laser fluence of 100 mJ/cm², the length of the remaining gold core is 53 nm and the volume 15852 nm³. Using the above formula, the diameter is coming out to be 19.6 nm.

In case of the laser fluence of 200 mJ/cm², the length of the remaining gold core is 50 nm and the volume 11889 nm³. Using the above formula, the diameter is coming out to be 17.4 nm.

Calculation of the used material starting from available Au and Pd in the nanorod:

We obtained an alloy composition of $Au_{1.51}Pd_{0.49}$, where the atomic ratio of gold to palladium is:

$$\text{Volumetric ratio} = \left(\frac{V_{Au}}{V_{Pd}}\right)_{\text{alloy}} = \frac{1.51}{0.49} \times \left(\frac{\text{Volume of Au unit cell}}{\text{Volume of Pd unit cell}}\right) = 7.09$$

The volume of gold used (at 100 mJ/cm²) in alloying is:

$$V_{Au, \text{used}} = 99077 \times 0.84 = 83224.68 \text{ nm}^3$$

The volume of palladium used (at 100 mJ/cm²) in the alloying process is:

$$V_{Pd, \text{used}} = \frac{83224.68}{7.09} = 11738.3 \text{ nm}^3$$

The volume of gold used (at 200 mJ/cm²) in alloying is:

$$V_{Au,\,used} = 99077 \times 0.88 = 87187.76 \text{ nm}^3$$

The volume of palladium used (at 200 mJ/cm²) in the alloying process is:

$$V_{Pd,\,used} = \frac{87187.76}{7.09} = 12297.3 \text{ nm}^3$$

---

### SI-6. Contribution of Debye-Waller factor[11,12] to diffraction intensity:

In SI-2 section, we observed that the achievable temperature of the gold nanorods at the used fluences of 10 mJ/cm² and 20 mJ/cm² are 670 K and 950 K, respectively. We aim to calculate the Debye-Waller factor at those temperatures and at 1333 K (melting temperature).

Basic Form:

$$I(T) = I_0 \, e^{-2W(T)}$$

Where:

I(T): Intensity at temperature T

$I_0$: Intensity at absolute zero (no vibration)

W: Debye-Waller parameter

$$W = \frac{1}{3} <u^2> Q^2$$

Where:

$<u^2>$: Mean square atomic displacement (thermal)

Q: Magnitude of scattering vector, $Q = \frac{4\pi \sin\theta}{\lambda}$

Calculation:

Under the high-temperature Debye approximation:

$$<u^2> = \frac{3\,\hbar^2 T}{M\,k_B\,\Theta_D^2}$$

Where:

$\hbar$: Reduced Planck constant = $1.0545 \times 10^{-34}$ J.s

$k_B$: Boltzmann constant = $1.380649 \times 10^{-23}$ J/K

T: Absolute temperature

M: Mass of the atom = $3.27 \times 10^{-25}$ kg

$\Theta_D$: Debye temperature of the material = 170 K

B-factor (temperature factor) which is related to the intensity by the following relation

$$I(T) = I_0 \, e^{-B \cdot Q^2/(4\pi^2)}$$

Where:

$$B = 8\pi^2 <u^2>$$

Combining the above information:

$$B = 8\pi^2 \frac{3\hbar^2 T}{M k_B \Theta_D^2}$$

$$I(T) = I_0 \cdot \exp\left(-\frac{8\pi^2 \cdot 3\hbar^2 T}{M k_B \Theta_D^2} \cdot \frac{Q^2}{4\pi^2}\right) = I_0 \cdot \exp\left(-\frac{6\hbar^2 T Q^2}{M k_B \Theta_D^2}\right)$$

At 670 K:

$$<u^2>_{670} = 0.0171$$

$$<u^2>_{298} = 0.0076$$

$$B_{670} = 1.352$$

$$B_{298} = 0.601$$

$$\frac{I_{670}}{I_{298}} = \exp\left(-\frac{(B_{670} - B_{298})}{4\pi^2} \cdot Q^2\right) = 0.8739$$

Relative intensity 87.39 %.

At 950 K:

$$<u^2>_{950} = 0.024$$

$$<u^2>_{298} = 0.0076$$

$$B_{950} = 1.917$$

$$B_{298} = 0.601$$

$$\frac{I_{950}}{I_{298}} = \exp\left(-\frac{(B_{950} - B_{298})}{4\pi^2} \cdot Q^2\right) = 0.7896$$

Relative intensity 78.96 %

At 1333 K:

$$<u^2>_{1333} = 0.034$$

$$<u^2>_{298} = 0.0076$$

$$B_{1333} = 2.69$$

$$B_{298} = 0.601$$

$$\frac{I_{1333}}{I_{298}} = \exp\left(-\frac{(B_{1333} - B_{298})}{4\pi^2} \cdot Q^2\right) = 0.6872$$

Relative intensity 68.72 %.

**SI-7. The amount of sample in a liquid jet interacting with an optical laser pulse:**

During the experiment, we employed an XFEL pulse structure operating at a frequency of 564

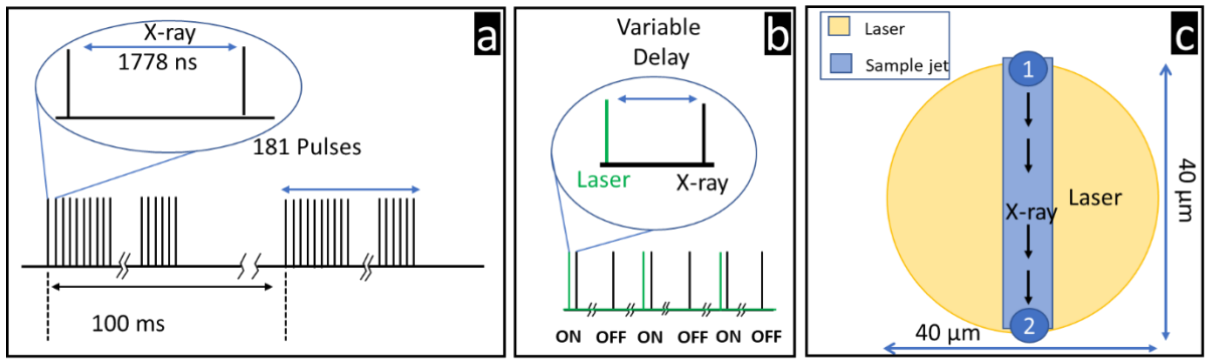

*Figure S10: (a) Schematic representation of the X-ray Free-Electron Laser (XFEL) pulse train structure. Each train consists of 181 X-ray pulses with an intra-train pulse-to-pulse separation of 1778 nanoseconds (ns), allowing time-resolved measurements across the full pulse train. (b) Optical laser ON-OFF scheme synchronized with the XFEL pulses. The optical laser operates at half the repetition rate of the XFEL, resulting in alternating laser-on and laser-off conditions for successive X-ray pulses to enable pump-probe differential measurements. (c) Schematic of the sample pumping and excitation geometry. The blue channel represents the flowing sample jet and X-ray beam size, while the yellow circular region indicates the laser excitation spot. The arrows show the direction of sample flow, with Point 1 marking the entry into the laser-irradiated region and Point 2 marking the exit.*

kHz with an inter-train repetition rate of 10 Hz. The intra-train pulse-to-pulse separation was 1778 ns, while the separation between pulse trains was 100 ms (Figure S10(a)). Optical laser pulses were used to excite the sample in a laser ON–OFF scheme, as illustrated in Figure S10(b), with precise synchronization between the optical laser and XFEL pulses. Prior to data collection, careful alignment ensured accurate spatial and temporal overlap among the optical laser, X-ray beam, and the sample stream. The velocity of particles within the liquid jet was measured to be approximately 40 m/s. Using this value, we calculated the transit time of nanoparticles across the ~40 μm diameter of the optical laser's focal spot. To guarantee single-pulse excitation, the optical laser repetition rate was set such that the pulse-to-pulse separation exceeded the nanoparticle's transit time through the interaction region. Furthermore, during laser ON–OFF measurements, we ensured that particles probed during OFF cycles had not been previously excited by adjusting the XFEL pulse spacing accordingly. This prevented residual excitation and ensured that dark-state data were collected from truly unpumped particles.

**SI-7.1. Limit of laser frequency increase:**

**Intra-train duty cycle:**

As mentioned above, we used the laser ON-OFF scheme for the measurement. The frequency of the optical laser pulses is 282 kHz, with a pulse-to-pulse separation of 3556 ns. Each train consists of 91 pulses. The Intra-train duty cycle is:

$$D_{Intra-train} = \frac{\text{laser pulse width}}{\text{pulse to pulse separation}} \times 100\%$$

$$= \frac{0.890 \text{ ps}}{(3.556 \times 10^6 \text{ ps})} \times 100\% = 2.5 \times 10^{-5} \%$$

**Inter-train duty cycle:**

The Inter-train duty cycle is:

$$D_{Inter-train} = \frac{\text{number of pulses in the train} \times \text{laser pulse width}}{\text{Pulse train separation}} \times 100\%$$

$$= \frac{91 \times 0.890 \text{ ps}}{10^{11} \text{ ps}} \times 100\%$$

$$= 8.09 \times 10^{-10} \%$$

So, in a continuous flow of the water jet we only excite $8.09 \times 10^{-10}$ % of the injected sample.

**Limit of laser frequency increase (repetition rate):**

The schematic shown in Figure S10(c) illustrates the flow of particles within the liquid jet, where the yellow circular region represents the optical laser spot with a diameter of 40 μm. The velocity of the liquid jet was measured to be approximately 40 m/s, a parameter that is relatively fixed and cannot be significantly altered. To ensure that each particle is exposed to the optical laser only once, the pulse-to-pulse separation of the optical laser was carefully selected. The transit time of a particle across the 40 μm diameter of the laser spot—from point 1 to point 2 is calculated as $(40 \times 10^{-6} \text{ m}) / (40 \text{ m/s}) = 1000$ ns. This transit time is shorter than both pulse-to-pulse separations of neighboring X-ray pulses (1778 ns) and neighboring optical laser pulses (3556 ns). As a result, there is no possibility of particles being excited more than once or of probing optically excited particles during the OFF cycle of the optical laser, ensuring clean separation between excited and dark-state measurements.

**SI-8. Comparative analysis of femtosecond laser-induced nanorod transformations:**

To contextualize the observed fluence threshold for alloying in our Au/Pd core–shell nanorods,

| Reference | Material System | Wavelength + Pulse Duration | Fluence (mJ/cm²)/ Pulse energy | Exposure Type | Laser Polarization | Laser Spot | Peak Intensity (W/cm²) | Observations |
|---|---|---|---|---|---|---|---|---|
| Zijlstra et al.,[13] | Au NRs (PVA matrix) | (700 nm – 1000 nm) 100 fs | 2.5 mJ/cm² | Single-shot | Linear | NA | $2.5\times10^{10}$ | Surface melting, spherical reshaping |
| Nazemi et al.,[14] | Au–Pd NRs (solution) | 808 nm 40 fs | ~ 3.6 mJ per pulse | Multi-shot (1000 Hz, 1 h) | Linear | NA | NA | Gradual Pd migration; no lattice alloying |
| Manzaneda-González et al.,[15] | Au@Pd Ag NRs | 800 nm 50 fs | 1 mJ/cm² ------------ 7 mJ/cm² | Multi-shot (1.5M pulses) | Linear | 10 mm -------- 6 mm | $2\times10^{10}$ ---------- $1.4\times10^{11}$ | Little change in aspect ratio at 1 mJ/cm² -------------- Hollowing and alloying at 7 mJ/cm² |
| González-Rubio et al.,[8] | Au NRs (solid) | 800 nm 50 fs | 0.32 mJ/cm² ---------- 0.51 mJ/cm² | Multi-shot | Linear | NA | ~$6.4\times10^{9}$ ------------ ~$1\times10^{10}$ | melting and symmetry loss in 60 min ------------ melting and symmetry loss in 10 min |
| Plech et al.,[16] | Au NPs | 400 nm 1 ps | 4 mJ/cm² | Single shot | Linear | 0.22 mm | ~$4\times10^{9}$ | Melting threshold |
| This Work | Au/Pd core–shell NRs (liquid jet) | 800 nm 890 fs | 100 mJ/cm², 200 mJ/cm² (measured) ------------ ~48 mJ/cm² (estimated melting) | Strict single-shot | Circular | 40 μm | $1.12\times10^{11}$ (for 100 mJ/cm²) ----------- $5.4\times10^{10}$ (for 48 mJ/cm²) | Lattice alloying inferred from TR-XRD; not major reshaping ---------- Melting |

*Table S2: Comparison of laser-induced transformations in nanorods.*

we compiled a comparison of representative literature studies involving femtosecond laser excitation of gold-based nanostructures. The key experimental parameters—material system, particle dimensions, excitation wavelength and pulse duration, polarization, exposure regime, and reported transformation mechanism—are summarized in Table S2. This overview highlights the broad range of reshaping and melting thresholds reported previously and illustrates that apparent discrepancies often arise from differences in particle composition, excitation conditions, and analysis approaches.

As seen in Table S2, reported fluence thresholds for reshaping or alloying are generally much lower than those observed here. However, most of those studies involved multi-pulse accumulation, shorter pulse durations with higher peak intensity, stationary or confined thermal environments, and compositionally softer nanostructures such as pure Au or alloyed shells. In contrast, our measurements probe single-pulse excitation of Au/Pd core–shell nanorods in a continuously refreshed liquid jet, where Pd shell stabilization, flow-induced alignment, circular polarization, and rapid heat dissipation all contribute to a different excitation regime.

| System | Particle type & size | Wavelength & pulse width | Reported threshold fluence | Absorption cross-section ($\sigma_{abs}$) | Atoms per particle (N) | Melting threshold energy per atom (eV/atom) | Reference |
|---|---|---|---|---|---|---|---|
| Au sphere | 44 nm diameter | 400 nm, 1 ps | 4 mJ/cm² | $3.9 \times 10^{-15}$ m² | ~ $2.6 \times 10^6$ | ~ 0.37 | Plech *et al.*,[16] |
| Au NRs | ~ 13 nm × 37 nm | 800 nm, 50 fs | 0.32 mJ/cm² ---------- 0.51 mJ/cm² | NA | ~ $8.2 \times 10^5$ | ~ 0.38 (reshaped less) ---------- ~ 0.6 (reshaped more) | González-Rubio *et al.*,[8] |
| Au/Pd core–shell NR | 106 nm × 44 nm | 800 nm, 890 fs | 48 mJ/cm² | $2.13 \times 10^{-15}$ m² | ~ $1.1 \times 10^7$ | ~ 0.57 | This work |

*Table S3: Comparison of upper energy gained by the atoms when illuminated by optical laser at threshold value.*

To provide a more quantitative comparison, we estimated the absorbed energy per atom at the reported threshold fluences, using the relation:

$$E_{abs} = \frac{F\, \sigma_{abs}}{N}$$

where F is the laser fluence, $\sigma_{abs}$ is the reported or estimated absorption cross-section, and N is the total number of atoms per nanoparticle. For our system, $\sigma_{abs}$ was derived directly from fluence-dependent lattice expansion measured by TR-XRD (see SI-3). The comparison across studies is summarized in Table S3.

Despite the order-of-magnitude difference in fluence thresholds (e.g., 4 mJ/cm² in Plech *et al.*,[16] 2024, vs. 48 mJ/cm² in this work), the absorbed energy per atom differs by only about a factor of 1.5 (0.37 vs. 0.57 eV/atom). This normalization shows that our results are energetically consistent with prior studies once differences in particle volume and atom count are taken into account.

More broadly, fluence alone is not a reliable metric for comparing nanoparticle transformation thresholds. The absorbed energy per atom depends simultaneously on the absorption cross-section, the number of atoms, and material- and shape-specific properties. Larger particles naturally contain more atoms, so even with similar cross-sections and fluences, the energy per atom is reduced. Core–shell architectures further modify both $\sigma_{abs}$ (through plasmon resonance shifts and damping) and the effective atom count (through shell thickness and density). Particle shape (spheres vs. rods) and material composition (Au vs. Pd vs. Ag) also strongly influence plasmonic coupling, cohesive energy, and diffusion kinetics. Finally, we note that earlier reshaping studies typically employed linearly polarized excitation, whereas our experiment used circular polarization. While we cannot definitively attribute the observed differences to this factor, the use of circular polarization might also contribute by reducing orientation-dependent excitation pathways that promote reshaping at lower fluences.

For these reasons, normalization by absorbed energy per atom, rather than fluence, provides a more meaningful basis for comparison. This analysis (Table S3) shows that the energies reached in our Au/Pd core–shell nanorods are compatible with the onset of alloying and melting, and explains why their fluence threshold appears higher while the per-atom energy is in line with prior reports.